\documentclass[10pt,a4paper]{article}
\usepackage[utf8]{inputenc}
\usepackage{amsmath, amsthm}
\usepackage{amsfonts}
\usepackage{amssymb}
\usepackage{bbm}
\usepackage{hyperref}
\usepackage{bm}
\usepackage{color}
\usepackage{float}
\usepackage{enumerate}
\usepackage[left=2cm,right=2cm,top=2cm,bottom=2cm]{geometry}
\addtocounter{tocdepth}{-2}
\usepackage{graphicx}
\usepackage{authblk}
\newcommand{\bbR}{\mathbb{R}}

\usepackage{epigraph}
\usepackage{float}

\title{The Strong Cosmic Censorship Conjecture}

\author[1]{Maxime~Van~de~Moortel\thanks{maxime.vandemoortel@rutgers.edu}}
\affil[1]{\small  Department of Mathematics, Rutgers University, 
	Hill~Center,~New~Brunswick~NJ~08854,~United~States~of~America \vskip.1pc \  }

\date{\today}

\usepackage[utf8]{inputenc}
\usepackage{amsmath, amsthm}
\usepackage{amsfonts}
\usepackage{amssymb}
\usepackage{bbm}
\usepackage{hyperref}

\begin{document}
	\maketitle
	\thispagestyle{empty}

	\begin{abstract} In the wake of major breakthroughs in General Relativity during the 1960s, Roger Penrose introduced Strong Cosmic Censorship, a profound conjecture regarding the deterministic nature of the theory. Penrose's proposal has since opened far-reaching new mathematical avenues, revealing connections to fundamental questions about black holes and the nature of gravitational singularities. We review recent advances arising from modern techniques in the theory of partial differential equations as applied to Strong Cosmic Censorship,    maintaining a focus on the context of gravitational collapse that gave birth to the conjecture.

	\end{abstract}

	\newtheorem*{rmk}{Remark}
	
	\numberwithin{equation}{section}
	\newtheorem{Theorem}{Theorem}

	\numberwithin{Theorem}{section}	
	\newtheorem{Def}[Theorem]{Definition}
	\newtheorem{Conjecture}[Theorem]{Conjecture}
	\newtheorem{Problem}[Theorem]{Problem}
	
	\maketitle
	
	\section{Introduction}\label{section.intro}

	\epigraph{The proving (or disproving)		of [...] Cosmic Censorship in classical general		relativity is perhaps the most important unsolved problem in that theory.}{Penrose (1982), \cite{Penrose1982}}
	\subsection{Cosmic Censorship: from the absence of naked singularities to  determinism}
	Penrose held the Cosmic Censorship hypothesis in  such high regard because, in his words, ``unlike many other problems'', it has ``genuine astrophysical significance'' \cite{Penrose1982}.   Cosmic Censorship is indeed inextricably tied to the very question of Laplacian determinism in the general theory of relativity; namely,  is ``the present state of the universe the effect of its past and  the cause of its future'' \cite{Laplace}? 
	Originally, the idea of Cosmic Censorship broadly asserts that should singularities exist within general relativity, they ought to be confined inside a black hole and, thus, do not affect the observable universe\footnote{Before the 1960's, relativistic singularities were viewed as absurdities that could, however, be ignored providing they were located inside a black hole, which was not then considered as a physically realistic spacetime, see \cite{Earman}, Chapter 3.2.}.		 	
	In other words, a cosmic censor  conceals any \emph{naked} (that is to say, visible to infinitely far-away observers) singularity within a black hole \cite{PenroseSCC}. 	However, very early on, Penrose motivated cosmic censorship as a way to avoid ``the unpredictability entailed by the presence of naked singularities'' \cite{Penrose1979}. In modern terms, the occurrence of  a naked singularity  is indeed responsible for the formation of a Cauchy horizon (see \cite{Hawking,ONeill} for a precise definition)  emanating from it. Should this Cauchy horizon be a regular hypersurface, the trajectory of observers crossing  it is no longer determined by their initial state. In Penrose's words: ``there being no theory governing what happens as the result of the appearance of such a singularity, this particular observer would not be able to account, in scientific terms, for whatever physical behaviour is seen'' \cite{Penrose1999}. In this situation, the future of spacetime itself beyond the Cauchy horizon cannot, in fact, be predicted from its initial state, in patent violation of the spirit of Laplacian determinism; such considerations will be discussed with greater depth in Section~\ref{section.hyp}.  \subsection{Locally naked singularities and   early statements of Strong Cosmic Censorship} Penrose  realized that  a  breakdown of Laplacian determinism would also occur if the naked singularity was covered by a black hole, an instance he calls ``a locally naked singularity'', namely a singularity that is still visible to observers falling into the black hole \cite{Penrose1979}.
	In his words, ``cosmic censorship [...] should  [..]  preclude such locally naked singularities'' \cite{Penrose1979}, a realization which	  prompted him to   introduce two distinct notions of cosmic censorship \cite{penrose1974gravitational}:\begin{itemize}
		\item  the absence of naked singularities in the process of gravitational collapse (the original cosmic censorship \cite{PenroseSCC}), dubbed  \emph{Weak Cosmic Censorship} by subsequent authors \cite{GerochHorowitz1979,ChristoCQG}.
		\item the absence of naked or locally naked singularities (possibly inside a black hole) in generic gravitational collapse, called by Penrose  ``\emph{Strong Cosmic Censorship}'' \cite{penrose1974gravitational,Penrose1979}.
	\end{itemize} The terminology is fully justified in these original forms because the  Strong Cosmic Censorship statement implies that of Weak Cosmic Censorship (see \cite{LandsmanCC}, section 5 for a historical perspective on this, but already contrast with the modern formulation of Strong Cosmic Censorship in our Section~\ref{modern.intro} below).   While originally, Cosmic Censorship concerned the  problem of naked singularities globally affecting the universe (especially idealized observers located at infinity), the change of paradigm in Penrose's work \cite{penrose1974gravitational,Penrose1979}  refocused Cosmic Censorship  on a failure  of Laplacian determinism, which can occur locally, with no reference to the observable universe in its globality.
	
	\subsection{The interior of the Kerr black hole and the  issue of Cauchy horizons} 
\begin{figure}[H]	\begin{center}\includegraphics[width=0.4\linewidth]{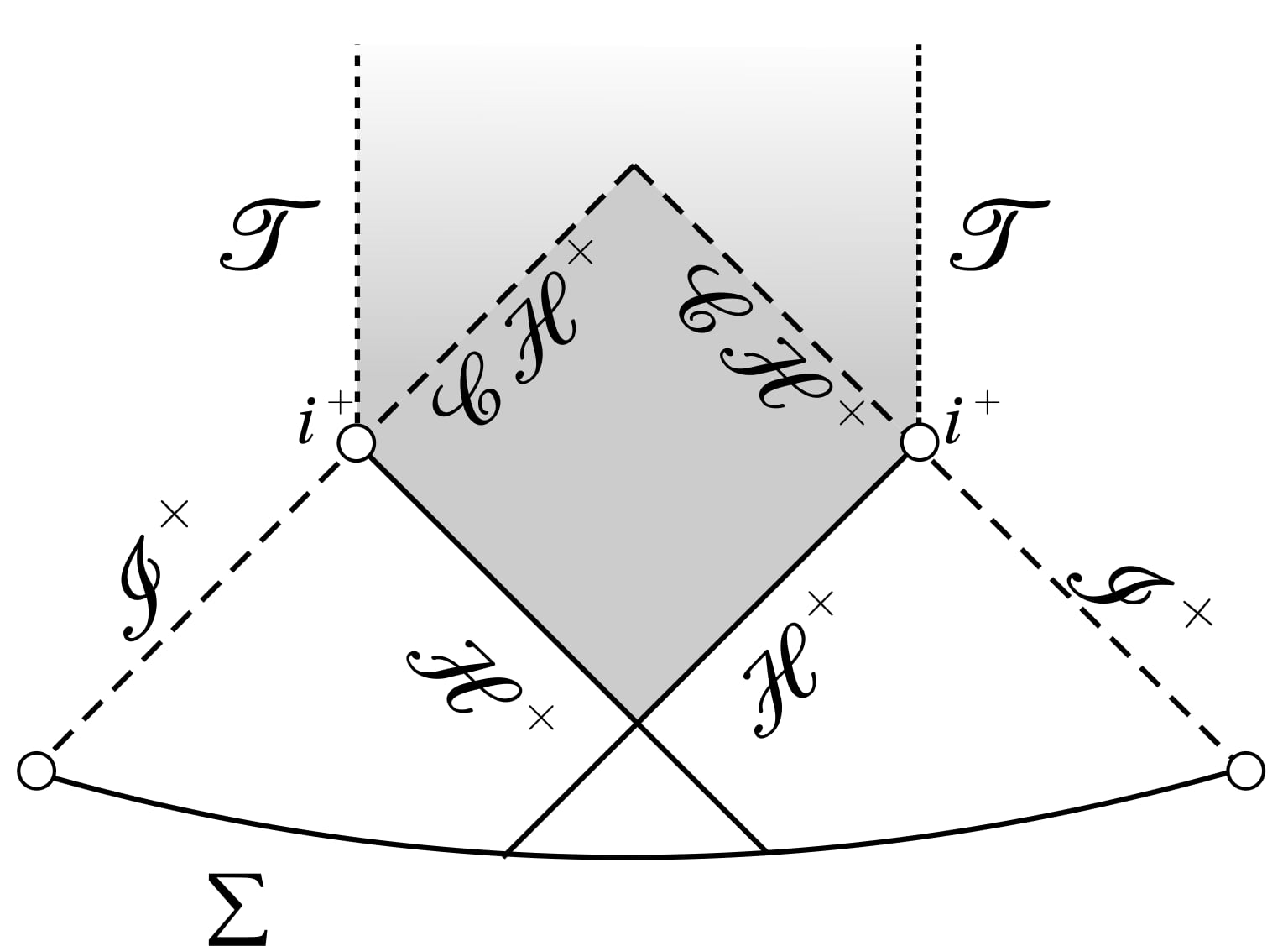}			\caption{\small Future analytic extension of the Kerr metric. $\mathcal{CH}^+$		 				 is a (smoothly extendible) Cauchy horizon emanating from timelike infinity $i^+$, and  $\mathcal{T}$ a timelike ring singularity.}		\label{fig:kerr}	\end{center}\end{figure}

The interior of the (sub-extremal) Kerr metric (representing, at the time of Penrose, the most appropriate known model for an astrophysical black hole), which also features a Cauchy horizon, provides a striking case study. Contrary to Cauchy horizons emanating from a naked singularity, the Kerr \emph{Cauchy horizon exists\footnote{The Kerr Cauchy horizon coincides with its inner horizon, and emanates from timelike infinity. This is another common mechanism for the appearance of Cauchy horizons, see \cite{Kommemi,MihalisSStrapped}.} despite the absence of any singularity} in its past. In retrospect, this fact demonstrates that the study of determinism (or its breakdown) is not directly tied to naked singularities and falls under more general considerations. That said, the Kerr metric interior (analytically extended to the future of its Cauchy horizon)  also features a timelike singularity, which is locally naked from the viewpoint of in-falling observers crossing the Cauchy horizon, see Figure~\ref{fig:kerr}. Penrose found such timelike singularities to be the most physically problematic issue with the Kerr metric (see \cite{LandsmanCC}, section 6). In \cite{Penroseblue,Penrose}, he  discovered that the Kerr Cauchy horizon suffers from a blue-shift instability 	   (we will return to this in Section~\ref{sectio.blue}). This instability mechanism gave credence to his original formulation of Strong Cosmic Censorship,  which precisely aimed at avoiding locally naked singularities. It is also important to note that the very existence of the Kerr metric  already demonstrates that
	Cauchy horizons (and the breakdown of  determinism they cause) cannot be  completely ruled out from General Relativity, which mandates a \emph{genericity condition} in the statement of Strong Cosmic Censorship (see Section~\ref{section.extend}), excluding the Kerr metric (see the related  discussion on blue-shift instability in Section~\ref{sectio.blue}).
	
	\normalsize

	\subsection{The modern statement of Strong Cosmic Censorship via global hyperbolicity}\label{modern.intro}

	Fifty years after its inception by Penrose,  the statement of Strong Cosmic Censorship remains true to its original spirit, while benefiting from the modern developments in the theory of the initial value problem for the Einstein equations. In 1974, Penrose already  proposed  a connection between Strong Cosmic Censorship and ``Leray's condition of global hyperbolicity'' \cite{penrose1974gravitational}. This insight later allowed for a clear formulation of Strong Cosmic Censorship in terms of the maximal globally hyperbolic development (MGHD), a concept introduced in the work of Choquet-Bruhat and Geroch \cite{GHD}. Given a suitable\footnote{By suitable initial hypersurface, it is meant a triplet $(M,g_0,K_0)$, where $(M,g_0)$ a complete three-dimensional Riemannian manifold, and $K_0$ a symmetric tensor such that $(M,g_0,K_0)$ satisfy the  Einstein constraint equations.} initial hypersurface (viewed as a snapshot of spacetime at time zero), its  MGHD represents the maximal spacetime that can be predicted by means of the Einstein equations from the  data posed on this initial hypersurface (see Section~\ref{subsection.MGHD}). 
	A  failure of determinism, however, arises if the MGHD is \emph{extendible as a solution to the Einstein equations}, for such an extension would be, by definition, not predicted  by the initial conditions generating the MGHD. In light of these connections,  Geroch and Horowitz \cite{GerochHorowitz1979} introduced the first modern   formulation of  Strong Cosmic Censorship as the impossibility of extending the MGHD for generic  asymptotically flat\footnote{Asymptotic flatness is  a natural assumption when studying  gravitational collapse, our main focus in this review. In the context of Cosmology, it is also interesting to study Strong Cosmic Censorship for compact initial data, see Section~\ref{section.variant}.} initial data. Such an approach \emph{excludes de facto timelike singularities} since they are incompatible with global hyperbolicity.		 Cauchy horizons, should they exist, are part of the future boundary of the MGHD  and, as such, the sole threat to Laplacian determinism by potentially allowing for spacetime extendibility. 
	The MGHD viewpoint leads to various similar formulations of Strong Cosmic Censorship \cite{ChristoCQG,MoncriefSCC,CIMSCC,Ringstrombook,nospacelike}, summarized in the following conjecture.
	\begin{Conjecture}[Strong Cosmic Censorship, modern formulation]\label{SCC.conj}
		For generic, asymptotically flat, complete, and regular initial data, the Maximal Globally Hyperbolic Development (MGHD) is inextendible as a solution to the Einstein equations.
	\end{Conjecture} The precise definitions of 'generic,' 'regular,' and 'solutions' are left intentionally vague to allow flexibility in solving the conjecture. It turns out that the validity of the conjecture varies according to the precise meaning of a ``solution'' to the Einstein equations (and its regularity): physically, this is  interpreted as the fact that determinism  only holds for spacetime metrics belonging to an adequate regularity class \cite{KerrStab,JonathanICM,ChristoSCC}; we  will come back to this in Section~\ref{subsection.MGHD}.

	\subsection{Modern Weak and Strong Cosmic Censorship, two independent conjectures}\label{subsection.independent}
	
	While the statement of Weak Cosmic Censorship, i.e.,  the absence of naked singularities  in generic gravitational collapse (see \cite{ChristoCQG,GerochHorowitz1979}), has remained essentially unchanged since its 1969 inception \cite{PenroseSCC}, the Strong Cosmic Censorship statement from Conjecture~\ref{SCC.conj} no longer explicitly refers to locally naked singularities contrary to Penrose's original version \cite{penrose1974gravitational}. As such,  the modern approach to both  Weak and Strong Cosmic Censorship renders both statements logically independent; in other words, \emph{Weak Cosmic Censorship does not follow from the resolution of Strong Cosmic Censorship}\footnote{In Christodoulou's words ``‘strong’ and ‘weak’ in reference to the two cosmic censorship
		conjectures is not a happy one, for it seems to suggest that the ‘strong’ conjecture implies the
		‘weak’ one, which is not the case.''  \cite{ChristoCQG}}.	 We note, however, that the presence of locally naked singularities inside black holes (irrelevant to Weak Cosmic Censorship in the form the conjecture is currently stated \cite{ChristoCQG,GerochHorowitz1979}) remains an obstruction to Strong Cosmic Censorship, both in its original formulation \cite{penrose1974gravitational,Penrose1979} and its modern form adopted in Conjecture~\ref{SCC.conj} (we will discuss this in more depth in Section~\ref{section.collapse}).

	\subsection{Strong Cosmic Censorship as a mathematical problem}  Beyond the fundamental issue of determinism posed by Strong Cosmic Censorship,  Conjecture~\ref{SCC.conj} turns out to be a source of profound mathematical questions at the crossroads of PDEs and Differential Geometry. A foundational problem in evolutionary PDEs is well-posedness:  does there exist a unique solution to the equations with prescribed initial data (existence/uniqueness)? Of course, in the context of the Einstein equations, one could interpret the existence of a unique MGHD as settling this question.  However, it is desirable to instead interpret the extendibility of the MGHD as \emph{a  failure of global uniqueness for the Einstein equations} -- an idea first formulated by Dafermos \cite{Mihalis1,Dafermos:2004jp,nospacelike} -- for there is no relativistic principle providing a  canonical extension of the MGHD -- should the MGHD be extendible: therefore, all possible extensions are arbitrary.
	
	We remark that, as  abstractly formulated in Conjecture~\ref{SCC.conj}, the ontology of Strong Cosmic Censorship is unrelated, prima facie, to singularities and black holes.
	However, it turns out that  an important obstruction to Strong Cosmic Censorship  precisely originates in the interior of black holes, as shown by the example of the Kerr metric that we now reexamine under a modern light. In  PDE terminology, the MGHD of Kerr initial data terminates at a Cauchy horizon, which is future-extendible and, thus, would represent a counter-example to Conjecture~\ref{SCC.conj} had we not included the word ``generic'' in its formulation (see Section~\ref{subsection.kerr}). It is precisely \emph{because} the MGHD of the Kerr metric is singularity-free (hence extendible) that, ironically, determinism  now appears  in jeopardy. Therefore, a restorationist strategy for determinism should involve showing  that for \emph{generic perturbations} of the Kerr black hole, the MGHD features a singular terminal boundary, across which no (future) extension is possible: this is indeed the hope entertained in the early works on blue-shift instability of Penrose \cite{Penroseblue} and Penrose--Simpson \cite{Penrose} that we discussed above.  More generally,  
	a resolution  of Conjecture~\ref{SCC.conj} would require showing the presence of a singularity in the interior of a generic black hole and thus answering the following fundamental  problem:\begin{Problem}\label{BH.problem}In the context of generic gravitational collapse, is\underline{} the terminal boundary of a black hole interior singular and, if so, how strong is the singularity?\end{Problem} The strength of the singularity is crucial, moreover, in that it precludes a failure of determinism in its corresponding regularity class \cite{KerrDaf,JonathanICM,MoiChristoph}. The theme of singularities in the context of black holes (and more broadly within general relativity) is indeed intimately connected to Cosmic Censorship; we will discuss this, together with  the celebrated singularity theorems  in Section~\ref{section.hyp}.	 		 
	
	Taking a step back, we may also connect such a mathematical endeavor with the more general theme of  \emph{dynamical singularity formation in evolutionary PDEs}, a class of differential equations to which the Einstein equations belong. 	 Finally, we emphasize that the connection between a singularity within the system of PDEs and the geometric inextendibility of the MGHD is subtle due to the covariant nature of the Einstein equations\footnote{the diffeomorphism-invariance of the theory indeed requires  to demonstrate the presence of a singularity in a  coordinate-independent fashion, which is a challenging task in general.},  which gives rise to fascinating geometric problems (see, in particular, Section~\ref{section.extend} and Section~\ref{sectio.blue}).

	We conclude this section  emphasizing	that  Strong Cosmic Censorship, despite its concise statement in Conjecture~\ref{SCC.conj}, encompasses  a web of mathematical problems and provides a testing field to rigorously understand  deep physical concepts such as singularities or determinism. The conjecture has nowadays become a field of research on its own and the source of many exciting mathematical developments. Progress toward its resolution is poised to have repercussions that extend far beyond  applications in General Relativity, particularly within the field of PDEs.

	\section{Initial value problem and Global Hyperbolicity}\label{section.hyp}
	
	\subsection{Laplacian determinism from Newtonian mechanics to General Relativity}\label{subsection.laplace}
	
	Newtonian mechanics has long been employed in the well-known classical theory of gravitation, the predecessor of General Relativity. In Newtonian mechanics, the motion of a point-particle of mass $m$ in a gravity field $\vec{g}$ is governed by Newton's second law  giving rise to the following ODE:
	
	\begin{equation}\label{newton}\begin{split}&  m \frac{d^2 \vec{x}}{dt^2} = m \vec{g}(x(t),y(t),z(t),t), \\ & \vec{x}(t=0)= \left(x_0,y_0,z_0\right),\   \frac{d\vec{x}}{dt}(t=0)= \left(\dot{x}_0,\dot{y}_0,\dot{z}_0\right).		\end{split}			\end{equation} 
	\eqref{newton} is par excellence deterministic in the sense of Laplace: indeed, for prescribed initial positions $\left(x_0,y_0,z_0\right)$ and velocities $\left(\dot{x}_0,\dot{y}_0,\dot{z}_0\right)$,  \eqref{newton} has a unique (possibly blowing-up in finite-time) solution. Hence, the initial state of the test particle fully determines its future evolution\footnote{For $N$ particles, however, \eqref{newton} becomes a system of nonlinear ODEs and its  solutions may blow-up in finite time (e.g.\ if two particles collide), which can cause a breakdown of determinism for this model if $N\geq 3$, see \cite{Earmanprimer}, Chapter III.}
	Analogously   in General Relativity, the motion of a freely falling test-particle (also known as a freely-falling \emph{observer}) on a Lorentzian manifold $(\mathcal{M},g)$ is governed by the timelike geodesic equation \begin{equation}\label{timegeodesic}\begin{split}
			&   \frac{d^2 \vec{x}^{\alpha}}{d\tau^2} +\Gamma^{\alpha}_{\sigma \mu}(g) \frac{d \vec{x}^{\sigma}}{d\tau} \frac{d \vec{x}^{\mu}}{d\tau}= \vec{0}, \\ & \vec{x}(\tau=0)= (x_0,y_0,z_0,t_0),\   \frac{d\vec{x}}{dt}(\tau=0)= (\dot{x}_0,\dot{y}_0,\dot{z}_0,\dot{t}_0),\\ & g( \frac{d\vec{x}}{dt}(\tau=0), \frac{d\vec{x}}{dt}(\tau=0))=-1.
		\end{split}
	\end{equation} where $\tau \geq 0$ represents the observer's proper time (note that $\vec{x}$ is now a four-dimensional spacetime vector) and $\Gamma^{\alpha}_{\sigma \mu}(g)$ are the Christoffel symbols of the spacetime metric $g$ (see e.g.\ \cite{ONeill,Hawking}).  
	\begin{rmk}A timelike geodesic  represents the wordline of a freely falling massive test-particle. The analogue for a massless test-particle (e.g.\ a photon) is provided by the null geodesic equation \begin{equation}\label{nullgeodesic}\begin{split}
				&   \frac{d^2 \vec{x}^{\alpha}}{d\tau^2} +\Gamma^{\alpha}_{\sigma \mu}(g) \frac{d \vec{x}^{\sigma}}{d\tau} \frac{d \vec{x}^{\mu}}{d\tau}= \vec{0}, \\ & \vec{x}(\tau=0)= (x_0,y_0,z_0,t_0),\   \frac{d\vec{x}}{dt}(\tau=0)= (\dot{x}_0,\dot{y}_0,\dot{z}_0,\dot{t}_0),\\ & g( \frac{d\vec{x}}{dt}(\tau=0), \frac{d\vec{x}}{dt}(\tau=0))=0.
			\end{split}
		\end{equation}
		Together, \eqref{timegeodesic} and \eqref{nullgeodesic} provide the equation of a  so-called causal geodesic.
	\end{rmk} 
	
	It is possible for a causal geodesic to exit the spacetime $(\mathcal{M},g)$ in finite proper time $\tau_{exit}$ (geodesic incompleteness, see Section~\ref{subsection.singthm}). In this situation, Laplacian determinism begs the question: do the equations predict what happens to this observer? We distinguish two possibilities:\begin{enumerate}[A.]
		\item \label{caseA} The observer encounters a terminal singularity, in the sense that \eqref{timegeodesic} cannot be continued beyond the exit proper time $\tau_{exit}$. This situation occurs for instance at the $\{r=0\}$ singularity in the interior of the Schwarzschild black hole (see Section~\ref{section.extend} for further details).
		\item \label{caseB} The observer does not encounter a terminal singularity, in the sense that $(\mathcal{M},g)$ can be (future) extended as a Lorentzian manifold in which \eqref{timegeodesic} makes sense for $\tau > \tau_{exit}$. This situation occurs  at the Cauchy horizon in the interior of the Kerr black hole, see Figure~\ref{fig:kerr}.
	\end{enumerate} 		
	A modern interpretation of  Scenario~\ref{caseA} is that Laplacian determinism is respected in this case \cite{Earman,MihalisLMS,JonathanICM,KerrStab}, highlighting the significance of singularities for Strong Cosmic Censorship. Scenario~\ref{caseB}, however, is not entirely conclusive; indeed, unveiling potential violations
	of determinism in this situation  additionally requires examining the predictability issues associated with equations governing the spacetime metric (e.g., Einstein equations), see Section~\ref{subsection.MGHD}. We note that the situation of   Scenario~\ref{caseB}  underscores the primacy of 
	\emph{spacetime extendibility} in resolving Conjecture~\ref{SCC.conj}, suggesting it is a more fundamental concept than the continuation of individual worldlines solutions of \eqref{timegeodesic} or \eqref{nullgeodesic}, as we discuss in Section~\ref{subsection.MGHD}.

	\subsection{Cauchy surfaces and global hyperbolicity}\label{subsection.cauchysurface}
	We now return to formulating the previously-described failure of determinism in terms of spacetime extendibility. We first describe general Lorentzian-geometric notions that do not require the Einstein equations \eqref{Einstein} to be satisfied, starting with the concept of a Cauchy surface. \begin{Def}[Cauchy surface] A hypersurface $\Sigma$ is a Cauchy surface for the Lorentzian manifold $(\mathcal{M},g)$ if it is met exactly once by any inextendible timelike  curve.
	\end{Def}
	The concept of a Cauchy surface is crucial in making sense of determinism within General Relativity for it provides an initial time hypersurface on which to prescribe initial conditions for the Einstein equations  \eqref{Einstein} (we will come back to this in Section~\ref{subsection.MGHD}). A spacetime in which any event is causally connected to such a Cauchy surface is called \emph{globally hyperbolic}. \begin{Def}[Globally hyperbolic spacetime] The Lorentzian manifold $(\mathcal{M},g)$ is called globally hyperbolic if it admits a Cauchy surface $\Sigma$.
	\end{Def} Note the region $\{t\geq 0\}$ on the maximal analytically-extended Schwarzschild spacetime (see Section~\ref{subsection.schw}) is globally hyperbolic, while the region $\{t\geq 0\}$ on the maximal analytically extended Kerr spacetime  is not \cite{Hawking,Wald}. Not every spacetime $(\mathcal{M},g)$ is globally hyperbolic, however, it is always possible \cite{ONeill} to construct a globally hyperbolic subset of $(\mathcal{M},g)$  by considering the (interior of the) domain of dependence $\mathcal{D}(\Sigma)$ of an achronal hypersurface $\Sigma$: in the case of $\Sigma=\{t=0\}$ in the Kerr spacetime, $\mathcal{D}(\Sigma)$ is the subregion of $\{t\geq0\}$ below the Cauchy horizon, see Figure~\ref{fig:kerr}.

	\subsection{The Einstein equations of spacetime dynamics and their initial value problem}\label{subsection.MGHD}~

	\textbf{The Einstein field equations.}	Contrary to Newtonian mechanics, the ontology of General Relativity is not about the trajectory of particles per s\'{e}, but instead about the dynamics of \emph{spacetime}, modeled by a four-dimensional Lorentzian metric $(\mathcal{M},g)$   governed by the Einstein  equations
	\begin{equation}\label{Einstein}
		Ric_{\mu \nu}(g)  -\frac{1}{2} Scal(g) g_{\mu \nu} = 8 \pi \mathbb{T}_{\mu \nu},
	\end{equation} where $Ric_{\mu \nu}(g)$ is the Ricci curvature tensor of the spacetime metric $g$ and $Scal(g)$ its scalar curvature. 	Here,  $\mathbb{T}_{\mu \nu}$  is the stress-energy tensor of the matter in presence, with $\mathbb{T}_{\mu \nu}\equiv 0$ for the Einstein vacuum equations. 
	
	\textbf{The Energy conditions.} The positivity of energy is reflected by the so-called energy conditions \cite{Hawking,Wald}, the weakest of which being the null energy condition for solutions of \eqref{Einstein}: \begin{equation}\label{null.energy}
		\mathbb{T}(v,v)\geq 0 \text{ for any null spacetime vector } v.
	\end{equation} Note  that \eqref{null.energy} can equivalently be  formulated purely in terms of the spacetime metric $g$ as $Ric(v,v) \geq 0$ for any null spacetime vector $v$, independently of \eqref{Einstein}. Analogously, one defines the strong energy condition  (which indeed implies the validity of the weak energy condition \eqref{null.energy}): \begin{equation}\label{time.energy}
		Ric(T,T)\geq 0 \text{ for any timelike spacetime vector } T.
	\end{equation}	
	
	\textbf{Classical solutions versus weak solutions.}	A classical solution $g$ of \eqref{Einstein} must possess twice differentiable components $g_{\mu \nu}$ in some appropriate coordinate system (dubbed a $C^2$ solution). Interestingly, however, it is possible to make sense of weak solutions of \eqref{Einstein} for continuous metrics $g$  with locally square-integrable Christoffel symbols \cite{ChristoSCC} (dubbed a $H^{1}_{loc}$ solution). These notions  play an important role in the formulation of Strong Cosmic Censorship, see  below. 
	
	\textbf{Initial value problem and Maximal Globally Hyperbolic Development.}	We are now ready to view the Einstein equations \eqref{Einstein} as an initial value problem analogously to \eqref{newton}. We call the following a suitable initial data set for the Einstein equations \eqref{Einstein} (say in vacuum, i.e., with $\mathbb{T}=0$). \begin{enumerate}
		\item A complete three-dimensional regular Riemmanian manifold $(\Sigma,g_0)$, which will later be viewed as the first fundamental form of the spacetime metric $g$ with respect to $\Sigma$.
		\item A symmetric regular two-tensor $K_0$, which will later be viewed as the second fundamental form of the spacetime metric $g$ with respect to $\Sigma$.
		\item  Assume $(g_0,K_0)$ satisfy the Einstein constraints equations induced by \eqref{Einstein} (see  e.g.\ \cite{Choquet-Bruhatbook}).
	\end{enumerate}
	
	To solve the initial value problem, one needs to treat $\Sigma$ as a Cauchy surface and construct a globally hyperbolic spacetime solution of \eqref{Einstein} consisting of the wordlines of timelike curves starting at $\Sigma$: the resulting Lorentzian manifold $(\mathcal{M},g)$ is the Globally Hyperbolic Development (GHD) of the initial data $(\Sigma,g_0,K_0)$, more precisely:
	
	\begin{Def} A globally hyperbolic development (GHD) $(\mathcal{M},g)$ of suitable initial data $(\Sigma,g_0,K_0)$ is a globally hyperbolic Lorentzian manifold $(\mathcal{M},g)$  satisfying \eqref{Einstein} as a classical solution 
		together					with an embedding  $i: \Sigma \rightarrow \mathcal{M}$ such that $i^{*}(g)=g_0$, $i^{*}(K)= K_0$, where $K$ is the second fundamental form of $i(\Sigma)$ in $\mathcal{M}$ with respect to the future normal and $i(\Sigma)$ is a Cauchy surface in $(\mathcal{M},g)$.
	\end{Def}
	The cornerstone of the initial value problem for the Einstein equations is given by the following result of Choquet-Bruhat and Geroch \cite{GHD} stating that there exists a unique maximal GHD of the prescribed initial data set $(\Sigma,g_0,K_0)$, considered  to be \emph{the} solution of \eqref{Einstein} for such initial data.
	
	\begin{Theorem}[\cite{GHD,dezorn}]\label{MGHD.thm} Given a suitable initial data set $(\Sigma,g_0,K_0)$, there exists a GHD $(\mathcal{M},g)$ that is an extension of any	other GHD of the same initial data. This GHD $(\mathcal{M},g)$ is unique up to isometry and is called the maximal globally
		hyperbolic development (MGHD) of the  initial data set $(\Sigma,g_0,K_0)$.
	\end{Theorem}
	\begin{rmk} In the above Theorem~\ref{MGHD.thm}, we  assumed regular  initial data; however, rougher solutions of \eqref{Einstein} are of considerable  mathematical and physical interest.  The state of the art is given  by the Klainerman--Rodnianski--Szeftel bounded-$L^2$ curvature theorem \cite{boundedL2}  proving the existence of a weak solution of \eqref{Einstein} in vacuum for initial data with two square-integrable derivatives (i.e., $H^2_{loc}$  in some coordinates). It is possible, however, to solve \eqref{Einstein} for even rougher initial data, provided they possess a special structure, for instance in the case of impulsive gravitational waves \cite{KhanPenrose,imp1,imp2,imp3paper1,imp3paper2}.
		
	\end{rmk}
	
	\textbf{Returning to Laplacian determinism: inextendibility of the MGHD.} The MGHD $(\mathcal{M},g)$ constructed in Theorem~\ref{MGHD.thm} consists of worldlines of inextendible timelike curves (i.e., observers). But what if the MGHD $(\mathcal{M},g)$ itself was extendible, i.e., there exists a Lorentzian manifold $(\bar{\mathcal{M}},\bar{g)}$ solution of \eqref{Einstein} and a proper embedding $ J: \mathcal{M}\rightarrow\bar{\mathcal{M}}$? Due to the hyperbolic nature of \eqref{Einstein}, one \emph{cannot} expect such an extension to be unique \cite{Dafermos:2004jp,KerrStab}, which already demonstrates a failure of global uniqueness for the initial value problem described above. This failure is also accompanied by a breakdown of Laplacian determinism, in the sense that the timelike curves which were inextendible in $(\mathcal{M},g)$ become extendible within  $(\bar{\mathcal{M}},\bar{g)}$: since the extension $(\bar{\mathcal{M}},\bar{g)}$ is arbitrary, so is the fate of these observers, which is not uniquely determined by their initial conditions at the Cauchy surface $\Sigma$. The validity of Laplacian determinism thus mandates the \textbf{inextendibility of the MGHD as a solution to the Einstein equations}, which motivated our formulation of Strong Cosmic Censorship as Conjecture~\ref{SCC.conj} in Section~\ref{modern.intro}. To conclude this section, we will formulate three physically-relevant notions of MGHD inextendibility of decreasing stength: \begin{enumerate}[I]
		\item\label{C0} ($C^0$ Strong Cosmic Censorship) The MGHD is inextendible as a $C^0$ Lorentzian manifold. This is the strongest possible inextendibility requirement at present, which  is a pointwise failure of continuity/boundedness  of the metric coefficients in \emph{every coordinate system}.
		\item\label{H1} ($H^1$ Strong Cosmic Censorship) The MGHD is inextendible as a $C^0$ Lorentzian manifold with square-integrable Christoffel symbols. Inextendibility at the $H^1$ regularity precludes any extension of the MGHD even as a weak solution to the Einstein equations \eqref{Einstein}.
		\item\label{C2} ($C^2$ Strong Cosmic Censorship) The MGHD is inextendible as a $C^2$ Lorentzian manifold with square-integrable Christoffel symbols. Inextendibility at the $C^2$ regularity precludes any extension of the MGHD as a classical solution to the Einstein equations \eqref{Einstein}.
	\end{enumerate}
	
	\begin{rmk}It might be possible, in principle,  to prove directly that the MGHD is inextendible as a (weak) solution to \eqref{Einstein}.  Instead, however, all known works on this question  prove inextendibility as a Lorentzian manifold within a given regularity class  such as the ones of Statement~\ref{C0}, \ref{H1}, or \ref{C2}.
	\end{rmk}				
	\subsection{Singularity theorems and the problem of geodesic incompleteness}\label{subsection.singthm}
	
	If the Strong Cosmic Censorship of Conjecture~\ref{SCC.conj} is true, then what is the mechanism leading to the inextendibility of the MGHD? In most cases, MGHD inextendibility is caused by the presence\footnote{Except if the MGHD is timelike geodesically complete, e.g.\ for Minkoski spacetime, in which case it is $C^0$ inextendible \cite{C0mink} despite the absence of singularities, see Section~\ref{subsection.mink} for a detailed discussion.} of a \textbf{singularity} on its the terminal boundary. The celebrated Penrose singularity  theorem \cite{Penrosesing} shows that the presence of a (closed) trapped surface, namely a topological sphere whose outgoing null mean curvature is negative (see e.g. \cite{claylecturenotes}), leads to geodesic incompleteness. 
	\begin{Theorem}[\cite{Penrosesing}]\label{penrose.sing.thm}If  $(\mathcal{M},g)$ 
		admits a non-compact				Cauchy hypersurface, contains a closed trapped surface
		and satisfies the  energy condition \eqref{null.energy}, then it is
		future-causally geodesically incomplete.
	\end{Theorem}  Note that the 
	singularity inside Schwarzschild's black hole (which, of course, contains a trapped surface) is indeed associated to causal geodesic incompleteness (see  Section~\ref{subsection.schw}). Because a surface being trapped is a stable condition with respect to small perturbations, Theorem~\ref{penrose.sing.thm} shows that geodesic incompleteness (interpreted as a form of singularity) ``is a robust prediction of the general theory of relativity'' \cite{penroseNobel}. However, it turns out that Theorem~\ref{penrose.sing.thm} merely asserts the existence of incomplete causal geodesics, which is not necessarily associated to a blow-up of curvature (or other gauge-invariant metric quantities): this is indeed what happens for the MGHD of the Kerr spacetime, when incomplete geodesics reach its Cauchy horizon as depicted in Figure~\ref{fig:kerr} \emph{despite the absence of any singularity at the Cauchy horizon} (Scenario~\ref{caseB} in Section~\ref{subsection.laplace}). For this reason, Theorem~\ref{penrose.sing.thm} is
	sometimes called the Penrose incompleteness theorem \cite{MihalisLMS,JonathanICM}.

	As a conclusion to this section, we emphasize that Conjecture~\ref{SCC.conj} requires to go beyond the statement provided by Theorem~\ref{penrose.sing.thm} and specify the nature of the geodesic incompleteness, particularly whether it is associated to a metric singularity strong enough to preclude extendibility in the sense of statements \ref{C0}, \ref{H1} or \ref{C2} described in the previous Section~\ref{subsection.MGHD}. Because trapped surfaces are the precursor of black holes, this mathematical endeavor is naturally connected to the fundamental Open Problem~\ref{BH.problem}, which will require addressing in any resolution of Conjecture~\ref{SCC.conj}.

	\section{Extendibility and singularity structure of particular solutions}\label{section.extend}
	While the extendibility properties of specific spacetimes technically has no bearing on Conjecture~\ref{SCC.conj}, since it only concerns \emph{generic solutions} of the Einstein equations \eqref{Einstein}, it is instructive to examine well-known examples in the light of the concepts developed in the earlier Section~\ref{section.hyp}.
	\subsection{Minkowski spacetime and other geodesically complete spacetimes}\label{subsection.mink} The Minkowski spacetime $(\bbR^{3+1},m)$ is the flat solution to \eqref{Einstein} in a vacuum given by \begin{equation}\label{minkowski}
		m= -dt^2 + dx^2+ dy^2+dz^2.
	\end{equation} $(\bbR^{3+1},m)$ is well-known to be causally geodesically complete, and stable to small perturbations from the celebrated work of Christodoulou--Klainerman \cite{stabmink} and Lindblad--Rodnianski \cite{stabminkharm}. Since both the Minkowski spacetime and its pertubations are singularity-free and causally geodesically complete, how do they fit in within the inextendibility requirements of Conjecture~\ref{SCC.conj}? As it turns out, any timelike geodesically complete spacetime (including $(\bbR^{3+1},m)$) satisfies the most demanding notion of Strong Cosmic Censorship, i.e., Statement~\ref{C0} in Section~\ref{subsection.MGHD}.
	
	\begin{Theorem}[\cite{C0mink}]\label{C0minkthm} A smooth (at least $C^2$) time-oriented Lorentzian manifold that is time-like geodesically complete and globally hyperbolic is $C^0$-inextendible.
	\end{Theorem}
	Therefore, Theorem~\ref{C0minkthm} shows that in proving Conjecture~\ref{SCC.conj}, one can now restrict our attention to timelike geodesically incomplete spacetimes, a large class of which is provided by spacetimes with a trapped surface (including any sub-extremal black holes), by Theorem~\ref{penrose.sing.thm}.
	
	\subsection{The Schwarzschild black hole}\label{subsection.schw}
	The Schwarzshild spacetime $(\bbR^{3+1},g_S)$ is  a one-parameter family of black hole solutions to \eqref{Einstein} in a vacuum indexed by their mass parameter $M>0$,  which takes for following form when $r\neq 2M$:\begin{equation}\label{schw}
		g_S= -(1-\frac{2M}{r})dt^2 + (1-\frac{2M}{r})^{-1}dr^2+ r^2 (d\theta^2+\sin^2(\theta) d\varphi^2).
	\end{equation} The spacetime geometry of the Schwarzshild spacetime $(\bbR^{3+1},g_S)$ is depicted in Figure~\ref{fig:schwarzschild}, with the event\footnote{The event horizon $\mathcal{H}^+=\{r=2M\}$ is a regular hypersurface for $g_S$ under the appropriate choice of coordinates \cite{Hawking,Wald}.} horizon $\mathcal{H}^+=\{r=2M\}$ and the spacelike singularity $\mathcal{S}=\{r=0\}$. The $\{r=0\}$ singularity is known to be terminal, in the sense that the Kretschmann scalar is infinite \cite{gravitation}  (curvature blow-up): it is thus well-known that the Schwarzschild spacetime is $C^2$-inextendible \cite{claylecturenotes} and, therefore, satisfies Strong Cosmic Censorship in the sense of Statement~\ref{C2} in Section~\ref{subsection.MGHD}. Moreover, it is also classically known \cite{gravitation,Wald} that in-falling observers reach  $\mathcal{S}=\{r=0\}$ in finite proper time and experience \emph{infinite tidal deformations}, which correspond, at least heuristically, to a metric blow-up on the $C^0$ level. These considerations beg the following question \cite{claylecturenotes}: is the Schwarzschild spacetime inextendible in a stronger sense, e.g. as in Statement~\ref{C0} of Section~\ref{subsection.MGHD}? The affirmative answer to this longstanding question was obtained by Sbierski \cite{JanC0} in the following theorem:
	\begin{figure}		\begin{center}\includegraphics[width=0.4\linewidth]{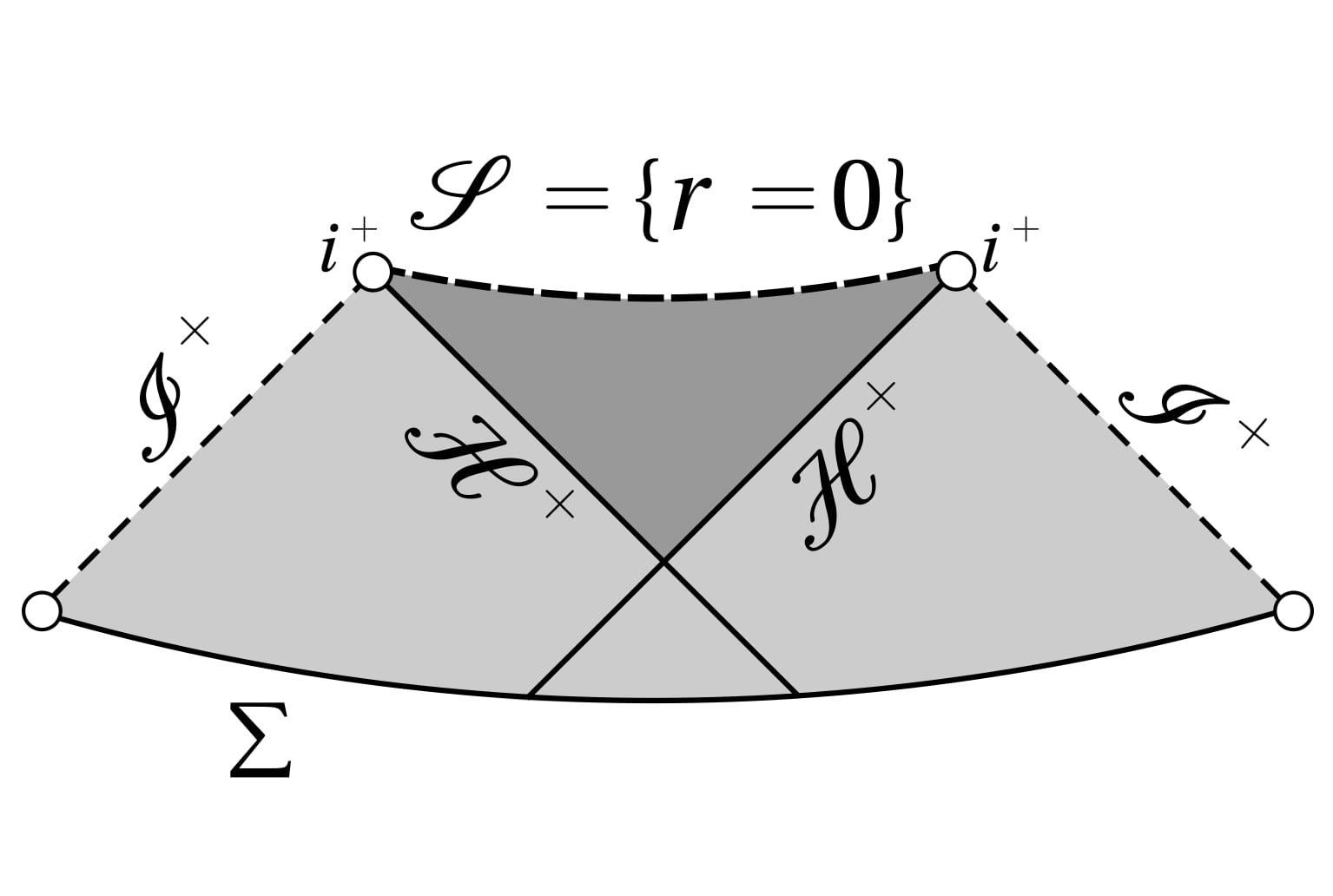}			\caption{\small Schwarzschild  as the MGHD of a two-ended asymptotically flat hypersurface $\Sigma$.\normalsize}		\label{fig:schwarzschild} \end{center}	\end{figure}
	\begin{Theorem}[\cite{JanC0}] \label{C0schwarzschild.thm} The MGHD of the (two-ended asymptotically flat) hypersurface $\{t=0\}$ of the Schwarzschild spacetime $g_S$ 
		is $C^0$-inextendible (to the future).
	\end{Theorem}

	\subsection{The Reissner--Nordstr\"{o}m and Kerr black holes} \label{subsection.kerr}The Schwarschild spacetime \eqref{schw} represents the simplest black hole, 
	uncharged and non-rotating. Its charged analogue is the Reissner--Nordstr\"{o}m spacetime $(\bbR^{3+1},g_{RN})$  of charge $e$ for $0< |e|\leq M$, which takes the following form when $r\neq r_{\pm}(M,e)$, where $r_{\pm}(M,e)= M\pm \sqrt{M^2-e^2}>0$:
	\begin{equation}\label{RN}
		g_{RN}= -(1-\frac{2M}{r}+\frac{e^2}{r^2})dt^2 + (1-\frac{2M}{r}+\frac{e^2}{r^2})^{-1}dr^2+ r^2 (d\theta^2+\sin^2(\theta) d\varphi^2).
	\end{equation}  and is a solution of the Einstein--Maxwell equations:
	\begin{equation}\label{EinsteinMaxwell}
		\begin{split}  Ric_{\mu \nu}(g)- \frac{1}{2}Scal(g)g_{\mu \nu}= 8\pi\underbrace{\left(g^{\alpha \beta}F _{\alpha \nu}F_{\beta \mu }-\frac{1}{4}F^{\alpha \beta}F_{\alpha \beta}g_{\mu \nu}\right)}_{=\mathbb{T}^{EM}_{\mu \nu}},\ \nabla^{\mu} F_{\mu \nu}=0.
		\end{split}
	\end{equation}
	
	Note that $r_{\pm}(M,e)$ correspond to the  roots of $(1-\frac{2M}{r}+\frac{e^2}{r^2})$, with $\mathcal{H}^+=\{r=r_+(M,e)\}$ being the event horizon  and $\mathcal{CH}^+=\{r=r_-(M,e)\}$  the Cauchy horizon: the MGHD of the Cauchy surface $\{t=0\}$ is $\{t\geq 0,\ r< r_-(M,e)\}$ and the Penrose diagram of the Reissner--Nordstr\"{o}m spacetime for $0<|e|<M$ is the same as that of Figure~\ref{fig:kerr}. The extremal Reissner--Nordstr\"{o}m spacetime corresponds to the case $|e|=M$.		The rotating analogue of the Schwarzschild spacetime is the celebrated Kerr \cite{MR0156674} black hole of angular momentum $0<|a|\leq M$ of the following form when $r\neq r_{\pm}(M,a)$:			
	\begin{equation}\label{Kerr}\begin{split}
			&	g_{K}=  - \frac{1-\frac{2M}{r}+ \frac{a^2}{r^2}}{1+\frac{a^2 \cos^2(\theta)}{r^2}}(dt-a \sin^2(\theta) d\varphi)^2+ \frac{1+\frac{a^2 \cos^2(\theta)}{r^2}}{1-\frac{2M}{r}+ \frac{a^2}{r^2}} dr^2\\ & +[r^2+a^2 \cos^2(\theta)] d\theta^2 +\frac{\sin^2(\theta)}{r^2+a^2 \cos^2(\theta)}[a dt- (r^2+a^2) d\varphi] ^2,\end{split}
	\end{equation}  and is a solution of \eqref{Einstein} in  vacuum ($\mathbb{T}=0$). Analogously, $\mathcal{H}^+=\{r=r_+(M,a)\}$ is the event horizon  and $\mathcal{CH}^+=\{r=r_-(M,a)\}$   the Cauchy horizon, where $r_{\pm}(M,a)= M\pm \sqrt{M^2-a^2}$; the MGHD is  $\{t\geq 0,\ r< r_-(M,a)\}$ and the Penrose diagram of the Kerr spacetime  for $0<|a|<M$ is given in Figure~\ref{fig:kerr}. Finally, the extremal Kerr spacetime corresponds to $|a|=M$.
	
	As already discussed in Section~\ref{section.intro}, the MGHD of the  Reissner--Nordstr\"{o}m and Kerr spacetimes are \emph{smoothly and non-uniquely extendible} across the Cauchy horizon $\mathcal{CH}^+$ and, therefore, do not even satisfy the least demanding inextendibility of Statement~\ref{C2}. This smooth extendibility makes for a serious threat to Laplacian determinism, all the more that the Kerr black hole is often considered to be a realistic model for astrophysical black holes. However, we emphasize that the  Reissner--Nordstr\"{o}m/Kerr spacetimes are not counter-examples to Strong Cosmic Censorship as formulated in Conjecture~\ref{SCC.conj}, precisely because they are \emph{not generic solutions} of \eqref{Einstein}, in view  of their  stationary character.  To prove Conjecture~\ref{SCC.conj}, one needs in particular prove the following: \begin{Problem}\label{Kerr.problem} Show that for  generic and smooth small perturbations  of the Reissner--Nordstr\"{o}m/Kerr initial data, the MGHD solving \eqref{Einstein} is inextendible as a suitably regular Lorentzian manifold.		
	\end{Problem} 				
	This genericity condition  may  appear as an all-too-convenient fiat in order to restore determinism. However, we emphasize that, in the Penrose blue-shift scenario \cite{Penroseblue,PenroseSCC}, the Reissner--Nordstr\"{o}m/Kerr interior region (i.e., the region $\{ r< r_+ \}$ beneath the event horizon) is conjecturally \emph{unstable to small perturbations}. Since a physically realistic solutions must be stable to small perturbations, the Reissner--Nordstr\"{o}m/Kerr interior region and their smooth extendibility can be deemed to be unphysical and therefore, it is legitimate to exclude them from the picture of gravitational collapse.
	That said, it is yet unclear  to which extent Penrose's blue-shift instability operates as a restorationist of determinism; we  will discuss this in more detail in Section~\ref{sectio.blue}.

	\subsection{The Oppenheimer--Snyder spacetime}\label{subsection.open}\begin{figure}[H]	\begin{center}\includegraphics[width=0.4\linewidth]{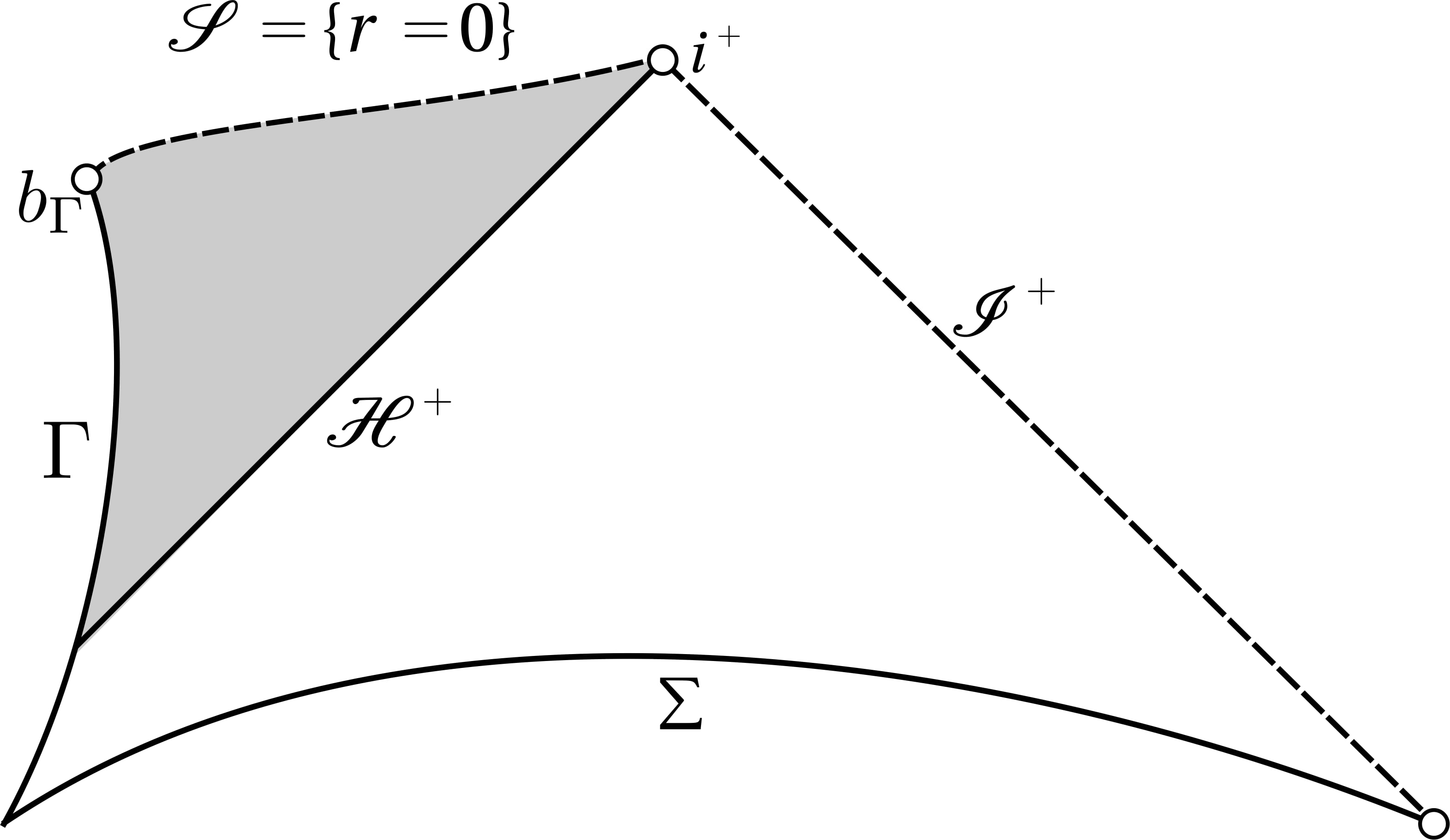}			\caption{\small Penrose diagram of  Oppenheimer--Snyder \cite{OppenheimerSnyder} spacetime (gravitational collapse).\normalsize}		\label{fig:collapse}	\end{center}\end{figure}
	
	The  first description of gravitational collapse came with the celebrated Oppenheimer--Snyder \cite{OppenheimerSnyder} black hole solution of \eqref{Einstein} with dust (a pressure-less perfect fluid with velocity $u^{\alpha}$, density $\rho$)  \begin{equation}
		\mathbb{T}_{\alpha \beta}^{dust}=\rho u_{\alpha}u_{\beta}.
	\end{equation}
	It consists of a ball of homogeneous dust of finite radius, outside of which the spacetime coincides with the Schwarzschild \eqref{schw} metric, and terminates at a spacelike singularity $\mathcal{S}=~\{r=~0\}$ as depicted in Figure~\ref{fig:collapse}. In view of Theorem~\ref{C0schwarzschild.thm}, we may conjecture that the Oppenheimer--Snyder spacetime is also $C^0$-inextendible. However, we emphasize that the Oppenheimer--Snyder spacetime does not represent a generic\footnote{In fact, even within the spherically-symmetric solutions of the Einstein-dust equations, the Oppenheimer--Snyder solution is non-generic, and some of its perturbations actually feature a naked singularity \cite{Christo4}! The standard  interpretation  is that dust is an unphysical matter model in these circumstances in which the fluid density becomes large \cite{Kommemi,claylecturenotes}.} instance of gravitational collapse, due to its high number of symmetries and the homogeneity of the dust ball, see Section~\ref{section.collapse} for further discussions.
	
	\subsection{The naked and locally naked singularities}\label{subsection.naked}
\begin{figure}		\begin{center}\includegraphics[width=0.8\linewidth]{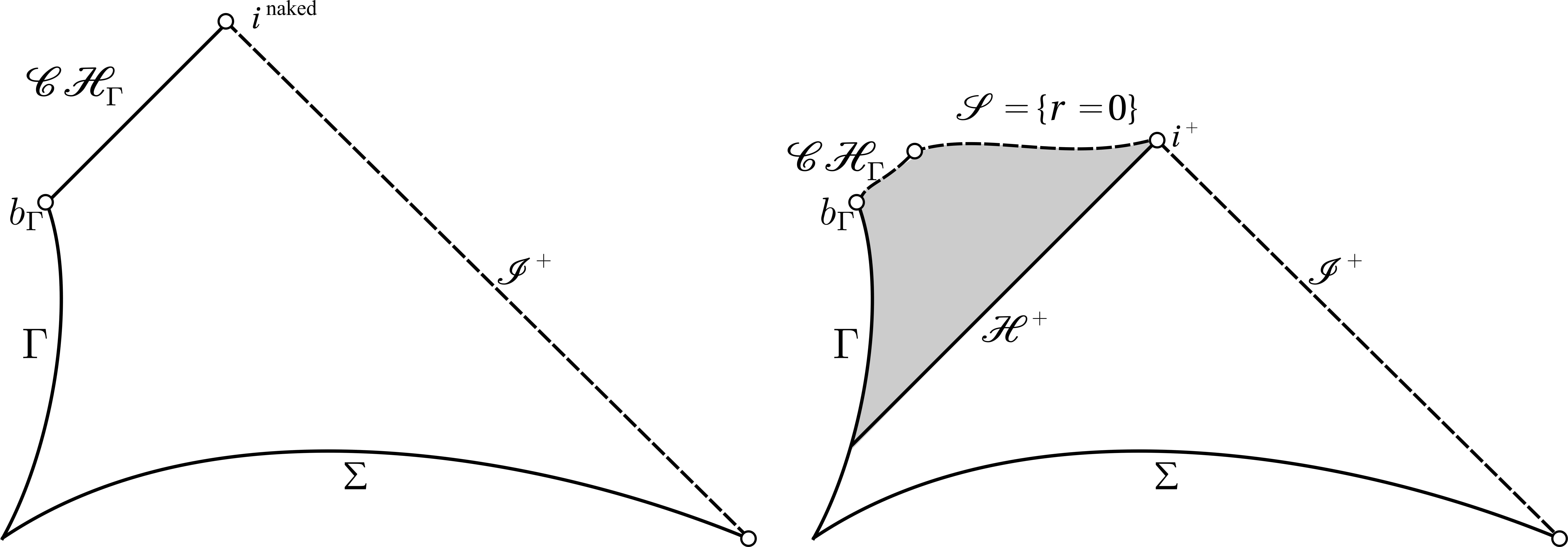}			\caption{\small Left: Penrose diagram of a naked singularity spacetime.\\ Right:   Penrose diagram of a spacetime with a locally naked singularity inside a black hole.\normalsize}		\label{fig:naked}	\end{center}\end{figure}

	In the seminal work \cite{Christo.existence}, Christodoulou constructed examples of naked singularities solutions of the Einstein-scalar-field equations, i.e., \eqref{Einstein} with the following scalar field stress-energy tensor \begin{equation}\label{ESF}
		\mathbb{T}_{\mu \nu}^{SF}= \partial_{\mu}\phi\partial_{\nu}\phi -\frac{1}{2}(g^{\alpha \beta} \partial_{\alpha}\phi \partial_{\beta}\phi  )g_{\mu \nu},\ \Box_g \phi =0,
	\end{equation} and moreover arising as a model of gravitational collapse (see Section~\ref{section.collapse}).							The Penrose diagram of Christodoulou's spacetime is depicted on the left of Figure~\ref{fig:naked}, and $b_{\Gamma}$ is a naked singularity.	Analogously \cite{Christo.existence} also constructs a \emph{locally naked singularity} $b_{\Gamma}$ inside a black hole region, see the right panel of Figure~\ref{fig:naked}: in both cases, a Cauchy horizon $\mathcal{CH}_{\Gamma}$ emanates from the singular point $b_{\Gamma}$ and across it, the spacetime is extendible\footnote{\label{foot}as a solution of the Einstein-scalar-field equations in the (rough) BV class defined by Christodoulou in \cite{Christo2}. In fact, the extension is $C^{1,\theta}$, which has the same regularity as the  initial data on a suitable Cauchy surface \cite{Christo3}.} as a solution to \eqref{Einstein}, \eqref{ESF}.
	The Christodoulou spacetimes  do not have smooth initial data (see  footnote~\ref{foot}) and therefore, do not, strictly speaking, pose a threat to Conjecture~\ref{SCC.conj}, which is formulated for solutions of \eqref{Einstein} with smooth initial data. The same is true for the recently-constructed naked singularity solutions of the vacuum \eqref{Einstein}   in the breakthrough of Shlapentokh-Rothman--Rodnianski \cite{nakedvacuum}.
	Within spherically symmetric matter models, naked singularities with \emph{smooth initial data}, however,  have been observed numerically \cite{Choptuik,Carsten}.  They pose a threat to Strong Cosmic Censorship, similarly to the  Cauchy horizons discussed  in Section~\ref{subsection.kerr}. In Section~\ref{section.collapse}, we will discuss strategies to show that such (locally) naked singularities are non-generic solutions, in order to prove Conjecture~\ref{SCC.conj}.
	
	\section{Cauchy horizons in dynamical black holes, blue-shift instability}\label{sectio.blue}
	
	\subsection{Blue-shift instability: heuristics and numerics} The celebrated blue-shift instability mechanism at the Reissner--Nordstr\"{o}m/Kerr Cauchy horizon is the cornerstone of Strong Cosmic Censorship as already envisioned by Penrose \cite{PenroseSCC}. 
	
\begin{figure}[H]\begin{center}\includegraphics[width=0.3\linewidth]{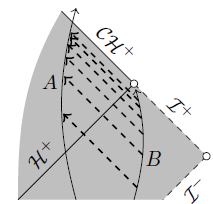}			\caption{\small Blue-shift effect at the Reissner--Nordstr\"{o}m/Kerr Cauchy horizon, figure from \cite{blueyakovM}.\normalsize}		\label{fig:blue}	\end{center}\end{figure}

	Consider two freely-falling  observers:  an in-falling timelike incomplete geodesic $A$ entering the black hole region,  reaching the Cauchy horizon $\mathcal{CH}^+$ in finite proper time, and    a complete timelike geodesic $B$ with infinite proper time  remaining outside of the black hole forever. The geometry depicted in Figure~\ref{fig:blue} is such that a pulse sent by $B$ towards $A$ with constant frequency becomes infinitely blue-shifted as $A$ reaches $\mathcal{CH}^+$. 	 The numerical study of linear fields on a fixed black hole (see  Section~\ref{subsection:linearblue}) confirms the blue-shift instability scenario, see \cite{bluenumerics,Penroseblue}.  How this blue-shift mechanism  impacts the Cauchy horizon stability for solutions of the Einstein equations \eqref{Einstein} (so-called back-reaction problem), however, remains to be seen, and will be discussed below.

	\subsection{Blue-shift in linear test-field theories on a fixed stationary black hole}\label{subsection:linearblue}
	
	Before turning to the full nonlinear dynamics of the Einstein equations \eqref{Einstein}, it is useful to first consider the behavior of linear test fields on a fixed  sub-extremal Reissner--Nordstr\"{o}m (i.e., \eqref{RN} with $0<|e|<M$)  or Kerr 		 (i.e., \eqref{Kerr} with $0<|a|<M$) 	black hole, starting with the wave equation\footnote{We  recall  that \eqref{waveBH} is often used as a linear toy model for the Einstein equations, due to their hyperbolic nature.} 						
	\begin{equation}\label{waveBH}
		\Box_{g} \phi =0,\ \text{ with } g=g_{RN} \text{ or }  g=g_{K}.
	\end{equation} The first theoretical  approach to establishing blow-up at the Cauchy horizon for solutions of \eqref{waveBH} (and their spin-weighted analogues, including electromagnetic and linearized gravitational fields) is due to McNamara 	\cite{McNamara} using a scattering method. Initial data are posed to be trivial at  the past event horizon $\mathcal{H}^{-}$  and a fast-decaying radiation field at past null infinity $\mathcal{I}^{-}$. The key idea, clarified in \cite{blueyakovM}, is to construct a \emph{transmission map} taking the initial data on $\mathcal{I}^-$ to the Cauchy horizon $\mathcal{CH}^+$, and use  the $t$-invariance of \eqref{RN} (or \eqref{Kerr}) to show that the local energy of the transmitted solution $\phi$ blows up at the Cauchy horizon $\mathcal{CH}^+$; In other words, $\phi \notin H^1_{loc}$, which is a linear analogue of the formulation \ref{H1} of Strong Cosmic Censorship from Section~\ref{subsection.MGHD}.	 The rigorous construction of the transmission map and its non-trivial character were proved in the work of Dafermos--Shlapentokh-Rothman \cite{blueyakovM}  establishing the following linear instability result\footnote{Anterior work of Sbierski \cite{beam} on Gaussian beams also provides solutions  with arbitrarily large local energy at $\mathcal{CH}^+$.}:
	\begin{Theorem} [\cite{blueyakovM}] There exists a 
		solution $\phi$ to \eqref{waveBH} such that $ r\phi_{|\mathcal{H}^{-}}\equiv 0$  and $ r\phi_{|\mathcal{I}^{-}}$ decays at an arbitrarily fast polynomial rate, yet the local energy of $\phi$ is infinite at the Cauchy horizon $\mathcal{CH}^+$.
	\end{Theorem}
	How to reconcile the existence of merely one solution with infinite local energy at the Cauchy horizon with the genericity requirement of Strong Cosmic Censorship? A well-known argument (see e.g. 	\cite{JonathanInstab}) shows that for \eqref{waveBH}, the existence of one blowing-up solution gives rise to blow-up for all initial data (in the same regularity class) except a co-dimension one subset: blow-up is thus generic in this sense. However, scattering-type blow-up arguments \`{a} la McNamara do not provide a constructive condition to decide whether a given solution to \eqref{waveBH} blows up at $\mathcal{CH}^+$ or not. A different approach to Cauchy horizon blow-up for \eqref{waveBH}, which relies on propagating inverse-polynomial lower bounds in time, has been proposed by Luk--Oh \cite{JonathanInstab} on the (sub-extremal) Reissner--Nordstr\"{o}m interior region \eqref{RN}, then by Luk--Sbierski \cite{KerrInstab} on  the (sub-extremal) Kerr interior   \eqref{Kerr}, starting from characteristic initial data on the event horizon $\mathcal{H}^+$. See also the more recent linear results for \eqref{waveBH} by Ma--Zheng \cite{Ma2} and Luk--Oh--Shlapentokh-Rothman \cite{massinflationYakov}.
	\begin{Theorem}[\cite{JonathanInstab,KerrInstab,Ma2,massinflationYakov}] \label{linear.int.thm}  If $\phi$ is a solution of \eqref{waveBH}  is such that its energy of   along the event
		horizon $\mathcal{H}^+$ obeys some polynomial upper and lower bounds then its local energy  is infinite at $\mathcal{CH}^+$.
	\end{Theorem}
	We also mention the recent work of Sbierski \cite{SbierskiTeukolsky} establishing an analogous result to Theorem~\ref{linear.int.thm} for the equations of linearized gravity; such results are poised to be used in an eventual proof of Strong Cosmic Censorship for the full (nonlinear) Einstein equations.		We finally  emphasize that the assumptions of Theorem~\ref{linear.int.thm} can be shown to be satisfied for \emph{generic} (sufficiently regular and localized) initial data at the Cauchy surface $\{t=0\}$ for \eqref{waveBH}, which becomes an asymptotic-in-time problem in the black hole exterior; we will return to this in Section~\ref{section.exterior}.

	\subsection{Nonlinear dynamics and spherically-symmetric matter models}\label{subsection.model}
	It is  natural to  consider Strong Cosmic Censorship for solutions to the Einstein equations in vacuum, as we will discuss below. 
	In order to understand delicate nonlinear issues like the ones associated to Strong Cosmic Censorship, it is useful to first look at spherically symmetric models, which are simpler to track. However, the well-known Birkhoff theorem (see e.g.\ \cite{Hawking,Wald}) shows that \emph{all spherically symmetric solutions of \eqref{Einstein} in vacuum are trivial}, in the sense that they are isometric to the Schwarzchild  \eqref{schw} or Minkowski metric \eqref{minkowski}. In other words, the spherically-symmetric Einstein equations  do not have a dynamical nature in vacuum  and are merely constraint equations. Therefore, to study  spherically symmetric dynamical problems in General Relativity, one must consider \eqref{Einstein} coupled to a non-trivial matter model $\mathbb{T}$. Ideally, the matter model should reintroduce the wave degrees of freedom lost by imposing spherical symmetry to the Einstein equations \cite{ChristoCQG}, which is what inspired Christodoulou to study the Einstein-scalar-field model \eqref{Einstein}, \eqref{ESF} in spherical symmetry in his series of works \cite{Christo1,Christo3,Christo.existence}. We will give a non-exhaustive list of such scalar-field matter models, in increasing order of sophistication.  
	\begin{itemize}

		\item (\textbf{Einstein-scalar-field}) The model \eqref{Einstein}, \eqref{ESF} was studied  by Christodoulou in \cite{Christo1,Christo2,Christo3} in his work on gravitational collapse  already mentioned in Section~\ref{subsection.naked}. Note that the Schwarzschild metric \eqref{schw} is a solution of \eqref{Einstein}, \eqref{ESF} with $\phi\equiv 0$. The Einstein-scalar-field spherically symmetric model does not allow for the presence of Cauchy horizons emanating from timelike infinity $i^+$, because of the absence of angular momentum or electric charge, so we will not  discuss it further in this section (see, however, Section~\ref{section.collapse}).  	\item (\textbf{Einstein--Maxwell-scalar-field}) To study Cauchy horizons while remaining in spherical symmetry, Dafermos \cite{Mihalis1,MihalisPHD} considered \eqref{Einstein} coupled with the  stress-energy tensor \begin{equation}\label{EMSF} \begin{split}	\mathbb{T}_{\mu \nu}=\mathbb{T}_{\mu \nu}^{EM}+\mathbb{T}_{\mu \nu}^{SF},\ \Box_g \phi =0,\ \nabla^{\mu} F_{\mu \nu}=0\end{split}				\end{equation} (recall $\mathbb{T}_{\mu \nu}^{EM}$ is defined in \eqref{EinsteinMaxwell}).  The Reissner--Nordstr\"{o}m metric \eqref{RN} is a solution of \eqref{Einstein}, \eqref{EMSF} with $\phi \equiv 0$.	Note that the uncharged scalar field $\phi$ is not directly coupled to the Maxwell field	$ F_{\mu \nu}$.		In the absence of charged matter, $ F_{\mu \nu}$ is subjected to a spherically-symmetric rigidity analogous to Birkhoff theorem \cite{Mihalis1,MihalisPHD,Kommemi} and takes the trivial form \begin{equation}\label{rigid.F} F= \frac{e}{r^2}\	 \Omega^2 du \wedge dv, \text{ where } g= -\Omega^2 du dv + r^2 d\sigma_{\mathbb{S}^2},																\end{equation} and $e\neq 0$ is a constant. Importantly, and contrary to Christodoulou's electromagnetism-free model, it is not possible to study gravitational collapse spacetimes within this model because such a situation requires the presence of a center $\Gamma=\{r=0\}$ (see Section~\ref{section.collapse}) where both the spacetime metric $g$ and the electromagnetic field $F$ are regular. The rigid form \eqref{rigid.F} and the  $e\neq 0$ are obviously incompatible with such a requirement at $r= 0$.
		\item (\textbf{Einstein--Maxwell-charged-scalar-field}) To remedy the Maxwell field rigidity of \eqref{rigid.F}, one can study a \emph{charged scalar field} \cite{Kommemi,Moi}, namely the  coupled stress-energy tensor: \begin{equation}\label{EMCSF} \begin{split} &	\mathbb{T}_{\mu \nu}=\mathbb{T}_{\mu \nu}^{EM}+\mathbb{T}_{\mu \nu}^{CSF},\ \mathbb{T}_{\mu \nu}^{CSF}=\Re(D_{\mu}\phi \overline{D_{\nu}\phi}) -\frac{1}{2}(g^{\alpha \beta} D_{\alpha}\phi \overline{D_{\beta}\phi}  )g_{\mu \nu},\\ & g^{\mu \nu} D_{\mu } D_{\nu} \phi =0,\ \nabla^{\mu} F_{\mu \nu}=i q_0 \Im(\bar{\phi} D_{\nu} \phi),\  D_{\mu} = \nabla_{\mu}+ i q_0 A_{\mu},\ F= dA,\end{split}				\end{equation} where $\nabla_{\mu}$ is the Levi-Civita connection of the spacetime metric $g$, $q_0 \neq 0$ is a coupling constant, and $A_{\mu}$ is the electromagnetic potential. Because the electromagnetic form $F$ is now coupled to the charged scalar field $\phi$, it is no longer subjected to the rigidity of \eqref{rigid.F}. Therefore, \eqref{EMCSF} is suitable as a spherically-symmetric model of gravitational collapse.
		
	\end{itemize} All the above spherically-symmetric models are generally considered as toy models for the Einstein vacuum equations (i.e., \eqref{Einstein} with $\mathbb{T}=0$). We emphasize that, strictly speaking, spherically-symmetric solutions of \eqref{Einstein} are non-generic and thus have no bearing to Conjecture~\ref{SCC.conj}; nonetheless, one can formulate an analogous simplified version of Conjecture~\ref{SCC.conj} within the class of spherically symmetric solution. 
	
	Finally, to complicate matters further, note that it is possible to study the \emph{massive Klein--Gordon} variant of all these models (see already Section~\ref{subsection.KG}), replacing $\Box_g \phi=0$ by \begin{equation}\label{KG}
		\Box_g \phi = m^2 \phi,
	\end{equation} and $m^2>0$. For instance, for an uncharged scalar field,  we have  the Einstein--Maxwell--Klein--Gordon equations: \begin{equation}\label{EMKG}
\mathbb{T}_{\mu \nu} =  \mathbb{T}_{\mu \nu}^{EM}+\mathbb{T}_{\mu \nu}^{KG},\ \Box_g \phi=m^2 \phi,\ \nabla^{\mu} F_{\mu \nu}=0,\ \mathbb{T}_{\mu \nu}^{KG} = \partial_{\mu}\phi\partial_{\nu}\phi-\frac{1}{2} [g^{\alpha \beta}\partial_{\alpha}\phi\partial_{\beta}\phi+m^2|\phi|^2] g_{\mu \nu}.
	\end{equation}

	The Einstein--(Maxwell)--Klein--Gordon equation, in contrast to the previously-mentioned models, is often viewed as a matter model in its own sake (as opposed to a toy model for vacuum dynamics)  which features different phenomena than the Einstein vacuum equations \cite{Yakov,KGSchw1,MoiChristoph,chodosh-sr1,chodosh-sr2}.

	\subsection{Einstein dynamics: nonlinear stability of the Cauchy horizon}\label{subsection:CHstab} One of the key issues in resolving Strong Cosmic Censorship is to understand the consequences of the blue-shift instability for dynamical perturbations of the  Reissner--Nordstr\"{o}m and Kerr spacetimes solving the Einstein equations \eqref{Einstein} either in vacuum or for a reasonable matter model. In view of the examples of Section~\ref{section.extend} (among the only examples known at the time of Penrose, see e.g. \cite{Hawking}), it might seem reasonable to speculate that the blue-shift instability ``destroys the Cauchy horizon'', and causes the dynamical appearance of a terminal singularity, the only known example of which being $\mathcal{S}=\{r=0\}$ in the Schwarzschild black hole \eqref{schw}. In the context of the vacuum Einstein equations, Penrose \cite{Penrose1979} writes `` there are some reasons for believing that
	generic perturbations
	away from spherical symmetry will not change the spacelike								nature of the singularity.'' Interpreting this scenario in the context of Conjecture~\ref{SCC.conj}, one is tempted to speculate about the validity Strong Cosmic Censorship in the $C^0$ sense, i.e., Statement~\ref{C0} from Section~\ref{subsection.MGHD} for small perturbations of the Kerr black hole, which leads to the following conjecture. \begin{Conjecture}[Spacelike singularity conjecture]\label{spacelike.conj} The MGHD of small perturbations of the Kerr metric \eqref{Kerr} solving the Einstein equations \eqref{Einstein} in a vacuum terminates at a spacelike singularity $\mathcal{S}$ and has the same Penrose diagram as Figure~\ref{fig:schwarzschild}. Moreover, the MGHD is $C^0$-future inextendible. \end{Conjecture}
	
	Dafermos studied this problem in spherical symmetry for solutions of the Einstein--Maxwell-scalar-field system \eqref{EMSF} converging to a sub-extremal  Reissner--Nordstr\"{o}m black hole, and \emph{surprisingly} proved that the Cauchy horizon ends up being stable to perturbations \cite{Mihalis1,MihalisPHD}, blue-shift instability notwithstanding! As a consequence, he also showed that the spacetime metric is continuously extendible, thus falsifying the spherically-symmetric version of Conjecture~\ref{spacelike.conj} and the $C^0$ version of Strong Cosmic Censorship given in Statement~\ref{C0} in Section~\ref{subsection.MGHD}. The author later \cite{Moi} established the Cauchy horizon stability for the more general Einstein--Maxwell-charged-scalar-field system \eqref{EMCSF} under which  the convergence to a sub-extremal  Reissner--Nordstr\"{o}m black hole is slower (see below), which necessitates the use of different methods from \cite{Mihalis1,MihalisPHD}. Below and in the rest of this review, we use the word \emph{``Cauchy horizon'' for a null component of the MGHD terminal boundary foliated by topological spheres  of non-zero volume} (see \cite{Kommemi,KerrStab,MihalisSStrapped}).
	
\begin{figure}[H]	\begin{center}\includegraphics[width=0.3\linewidth]{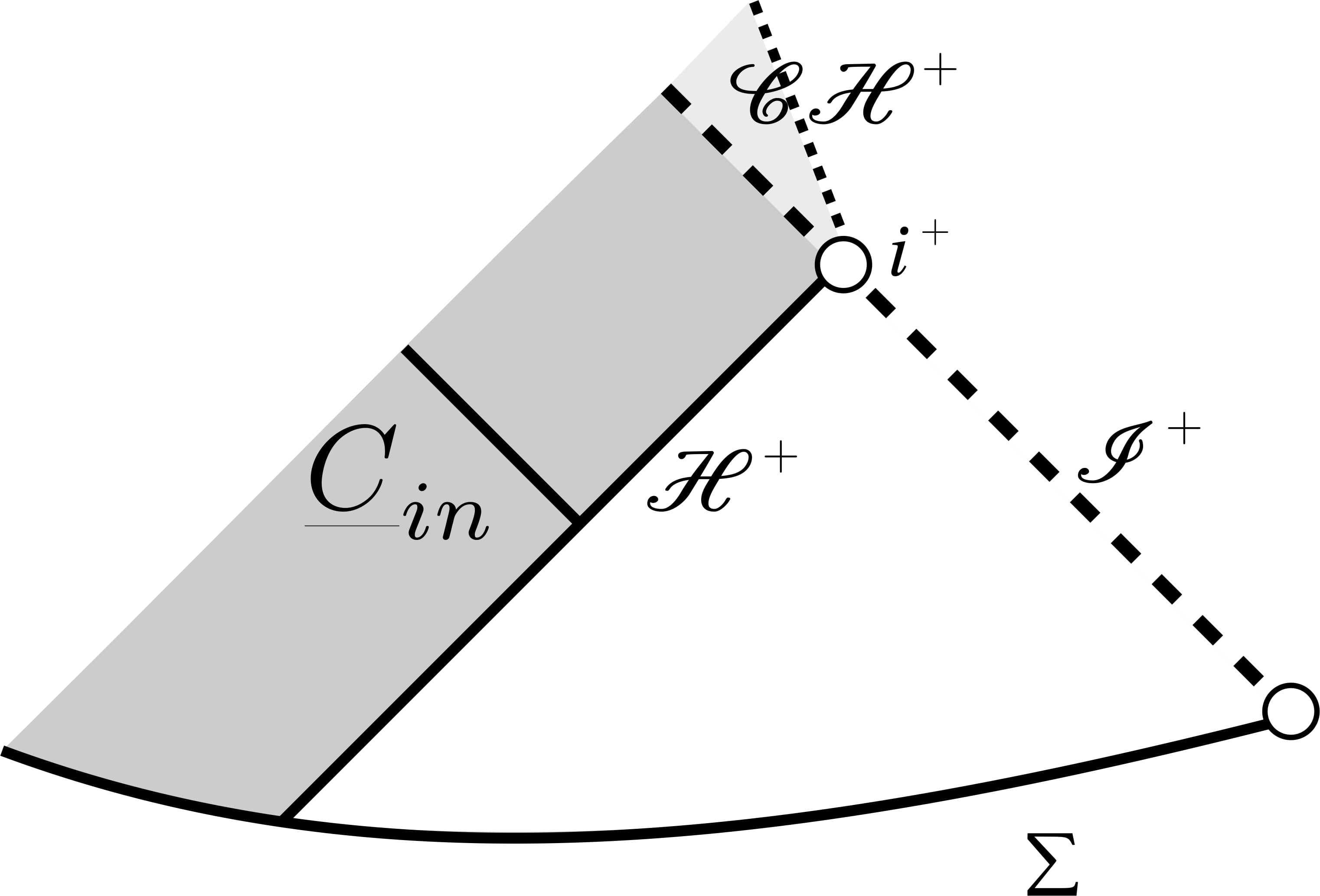}			\caption{\small Characteristic initial data in the black hole interior and continuous extension across the Cauchy horizon $\mathcal{CH}^+$ for Theorem~\ref{CH.thm.SS}, Theorem~\ref{CH.osc.thm}, Theorem~\ref{CH.thm.vacuum}.}	\end{center}	\label{fig:interior}	\end{figure}

	\begin{Theorem}\label{CH.thm.SS}Consider spherically symmetric characteristic initial data on $\underline{C}_{in} \cup \mathcal{H}^+$ as depicted in Figure~\ref{fig:interior}, and under the assumption that $(g,\phi)$ converge to a sub-extremal Reissner--Nordstr\"{o}m black hole with  inverse-polynomial upper bounds  for $\phi$ on the event horizon $\mathcal{H}^+$.
		\begin{itemize}
			\item \cite{Mihalis1,MihalisPHD,JonathanStab} For the Einstein--Maxwell-scalar-field system \eqref{EMSF}, the solution admits a non-empty Cauchy horizon $\mathcal{CH}^{+}$ across which the metric $g$ is continuously extendible. 
			\item \cite{Moi}  For the Einstein--Maxwell-charged-scalar-field system \eqref{EMCSF}, the solution admits a non-empty Cauchy horizon $\mathcal{CH}^{+}$. 
		\end{itemize}
		
	\end{Theorem}	

	The question of continuous extendibility for the Einstein--Maxwell-charged-scalar-field system \eqref{EMCSF} is left open in Theorem~\ref{CH.thm.SS}; it is directly related to the pointwise boundedness of the scalar field at the Cauchy horizon, i.e., is it true that $\phi\in L^{\infty}$? Because the decay assumptions on the event horizon $\mathcal{H}^+$ are weaker than in the uncharged-scalar-field case, this issue becomes delicate and depends on the exact profile of the characteristic initial data $\phi_{|\mathcal{H}^+}$ (see already Theorem~\ref{CH.osc.thm} below). More precisely, the decay assumptions in Theorem~\ref{CH.thm.SS} take the following form in an Eddington--Finkelstein advanced-time coordinate $v$ on $\mathcal{H}^+$ as $v\rightarrow +\infty$,  for some $s>\frac{3}{4}$:
	\begin{equation}\label{decay}
		[1+|v|^{s}]  \phi_{|\mathcal{H}^+}(v) \text{ and its derivative are bounded in an unweighted-in-}v\text{ norm.} 
	\end{equation}
	For the	Einstein--Maxwell-scalar-field system \eqref{EMSF}, one can prove \eqref{decay}  for $s=3$ for spherically-symmetric solutions with regular and localized initial data \cite{PriceLaw}, while for the	Einstein--Maxwell-charged-scalar-field  \eqref{EMCSF}, the conjecture \cite{MoiChristoph} is that for black holes of asymptotic charge $e\neq 0$ \begin{equation}\label{charged.decay.conj}
		s(q_0e)= 1+\sqrt{1-4 (q_0 e)^2} \text{ if } |q_0 e|<\frac{1}{2},\ s(q_0 e)=1 \text{ otherwise}.
	\end{equation} (we will discuss these decay rates  in Section~\ref{section.exterior}). In Theorem~\ref{CH.thm.SS}, continuous extendibility can be proven\footnote{For $s>1$, continuous extendibility is  deduced from a  stronger  extendibility property of $(g,\phi)$ in a $W^{1,1}$-type norm.} provided \eqref{decay} with $s>1$ is assumed, which only makes sense for \eqref{EMSF} or \eqref{EMCSF} if $|q_0 e|<\frac{1}{2}$. 
	However, under the general assumptions of Theorem~\ref{CH.thm.SS}, some solutions are $C^0$-extendible, but  others are not \cite{MoiChristoph}, depending on the scalar field  oscillations   on $\mathcal{H}^+$: the author  proved with Kehle that

	\begin{Theorem}[\cite{MoiChristoph}]\label{CH.osc.thm} Under the same assumptions as Theorem~\ref{CH.thm.SS} for the Einstein--Maxwell-charged-scalar-field \eqref{EMCSF}, assume additionally that										 $\phi_{|\mathcal{H}^+}$ oscillates sufficiently. 
		Then $\phi \in L^{\infty}$ and $(g,\phi)$ are continuously extendible across the Cauchy horizon $\mathcal{CH}^+$.
	\end{Theorem} The oscillation condition of Theorem~\ref{CH.osc.thm} involves the finiteness of integrals of the form \begin{equation}\label{osc}
		\bigl|\int_{1}^{\infty} \phi_{|\mathcal{H}^+}(v) e^{-i[\omega_{res}+o(1)] v}dv\bigr| < \infty,
	\end{equation}where $\omega_{res}=q_0 e[\frac{1}{r_+}-\frac{1}{r_-}]\in \bbR$; \eqref{osc} is a semi-integrability condition  in particular satisfied if \eqref{decay} holds for $s>1$, but not necessarily otherwise.				The oscillation condition \eqref{osc} is conjectured to be satisfied \cite{HodPiran1,HodPiran2} for generic solutions of \eqref{Einstein}, \eqref{EMCSF}, see  the discussion in Section~\ref{section.exterior}.

	We now return to the vacuum Einstein equations \eqref{Einstein} and discuss the breakthrough of Dafermos--Luk \cite{KerrStab} proving that vacuum black holes converging to Kerr \eqref{Kerr} also admit a $C^0$-extendible Cauchy horizon, and thus offering a definitive disproof of Conjecture~\ref{spacelike.conj}.
	\begin{Theorem}[\cite{KerrStab}]\label{CH.thm.vacuum} Consider  characteristic initial data on $\underline{C}_{in} \cup \mathcal{H}^+$ as depicted in Figure~\ref{fig:interior}, and under the assumption that $g$ converge to a sub-extremal Kerr black hole with inverse-polynomial upper bounds  on the event horizon $\mathcal{H}^+$. Then, the resulting solution $(\mathcal{M},g)$ of \eqref{Einstein} in vacuum  admits a non-empty Cauchy horizon $\mathcal{CH}^{+}$ across which the metric $g$ is continuously extendible. 
	\end{Theorem}
	Together with the stability of the Kerr black hole exterior proven in	\cite{klainerman2021kerr,GKS} (see Section~\ref{section.exterior}), Theorem~\ref{CH.thm.vacuum} also falsifies the $C^0$-version of Strong Cosmic Censorship (statement~\ref{C0} in Section~\ref{subsection.MGHD}).   However, despite Conjecture~\ref{spacelike.conj} being false, it is still possible that during gravitational collapse, \emph{part of the terminal boundary is spacelike}; this will a major topic discussed in Section~\ref{section.collapse}.
	
	\subsection{Einstein dynamics: nonlinear instability of the Cauchy horizon, mass inflation}
	We now return to Open Problem~\ref{Kerr.problem}, which is an essential piece in a potential resolution of Strong Cosmic Censorship.
	The Reissner--Nordstr\"{o}m/Kerr Cauchy horizons dynamical stability obtained in Section~\ref{subsection:CHstab} begs the question: what are the consequences of the blue-shift instability discussed in Section~\ref{subsection:linearblue} in the nonlinear dynamics picture? The falsification  of Conjecture~\ref{spacelike.conj} already shows that the blue-shift instability \emph{does not destroy the Cauchy horizon}, which remains continuously extendible. Could it then be that it is also smoothly extendible as in the Reissner--Nordstr\"{o}m/Kerr cases discussed in Section~\ref{section.extend}? The key insight comes from the mass inflation scenario of Poisson--Israel \cite{Poisson,Poisson2} and Ori \cite{Ori} (see also their precursor \cite{Hiscock}). They study \eqref{Einstein} in spherical symmetry coupled to a  rudimentary matter model in which two clouds of dust are transported in the ingoing and outgoing null directions. Their analysis involves the Hawking mass $\rho$, defined in terms of the area-radius function $r$ of the spherically-symmetric  metric $g$ as\begin{equation}\label{rho.def}
		\rho:= \frac{r[1-g(\nabla r,\nabla r)]}{2}.
	\end{equation} They showed that, for non-trivial ingoing and outgoing event-horizon radiation  falling-off at an inverse polynomial rate as in the theorems of Section~\ref{subsection:CHstab}, the Hawking mass $\rho$ and the Riemann curvature blow up at the Cauchy horizon. In summary, Strong Cosmic Censorship, at least in the weaker sense of Statement~\ref{H1} and Statement~\ref{C2} from Section~\ref{subsection.MGHD}, can still be valid \emph{despite the existence of Cauchy horizon} in dynamical spacetimes. For the Einstein--Maxwell-scalar-field system \eqref{EMSF}, Dafermos established the blow-up of  $\rho$ under the inverse-polynomial decay assumption \eqref{decay} together with an associated pointwise lower bound assumption \cite{Mihalis1,MihalisPHD}. The blow-up of curvature was established by Luk--Oh  \cite{JonathanStab} under a more relaxed inverse-polynomial lower bound assumption on the energy of $\phi_{|\mathcal{H}^+}$, while the blow-up of $\rho$ was obtained under similar assumptions in \cite{massinflationYakov}.  Curvature blow-up for the charged scalar field case \eqref{EMCSF} was obtained in the author's work \cite{Moi,Moi4}. All these results are summarized in the following theorem.
	\begin{Theorem}\label{CH.instab.thm.SS} In the setting and under the same assumptions as Theorem~\ref{CH.thm.SS}, assume additionally that \eqref{decay} also holds as a lower bound on the event horizon $\mathcal{H}^+$ in a suitable norm.
		\begin{itemize}
			\item \cite{Mihalis1,MihalisPHD,JonathanStab,massinflationYakov} For the Einstein--Maxwell-scalar-field system \eqref{EMSF}, the  Cauchy horizon $\mathcal{CH}^{+}$ is weakly singular, in the sense that $\rho$ and certain curvature components blow-up.
			\item \cite{Moi,Moi4}  For the Einstein--Maxwell-charged-scalar-field system \eqref{EMCSF},  the  Cauchy horizon $\mathcal{CH}^{+}$ is weakly singular, in the sense that  certain curvature components blow-up. 
		\end{itemize}
		
	\end{Theorem}
	
	We emphasize that lower bounds on the scalar field are  essential in Theorem~\ref{CH.instab.thm.SS}, since when $\phi\equiv 0$ (Reissner--Nordstr\"{o}m case), the Cauchy horizon $\mathcal{CH}^{+}$ is not singular (recalling Section~\ref{section.extend}). In Section~\ref{section.exterior}, we will see that these lower bounds on the event horizon $\mathcal{H}^+$ are satisfied for spherically symmetric solutions of \eqref{EMSF} with generic, localized and regular Cauchy data in  the black hole exterior \cite{JonathanStabExt}, thus connecting  with  Conjecture~\ref{SCC.conj}  restricted to spherical symmetry.
	
	In terms of inextendibility, Theorem~\ref{CH.instab.thm.SS} leads to the proof that there \emph{does not exist} any $C^2$ extension of the metric \cite{JonathanStab,Moi,Moi4} across $\mathcal{CH}^+$, consistently with Statement~\ref{C2} in Section~\ref{subsection.MGHD}.
	
	Finally, we return to the Einstein vacuum equations (\eqref{Einstein}  with $\mathbb{T}=0$) for which the first examples of a black hole with a weakly singular Cauchy horizons were constructed by Luk \cite{JonathanWeakNull}. \begin{Theorem}[\cite{JonathanWeakNull}]\label{weaknull.thm} There exists a class of  black hole interior  solutions of the Einstein vacuum equation									 without any symmetry assumptions featuring a weakly singular Cauchy horizon $\mathcal{CH}^+$.\end{Theorem}
	The stability estimates in 		Theorem~\ref{CH.thm.vacuum} are consistent with Theorem~\ref{weaknull.thm} and suggest that  a generic perturbation of the Kerr black hole features a weak singularity similar to the constructions of \cite{JonathanWeakNull}.
	The analogous result to Theorem~\ref{CH.instab.thm.SS} for \eqref{Einstein} in vacuum, however, is still open at present.

	\subsection{Strong Cosmic Censorship for  two-ended spacetimes}\label{subsection.two} 
\begin{figure}	\includegraphics[width=0.82\linewidth]{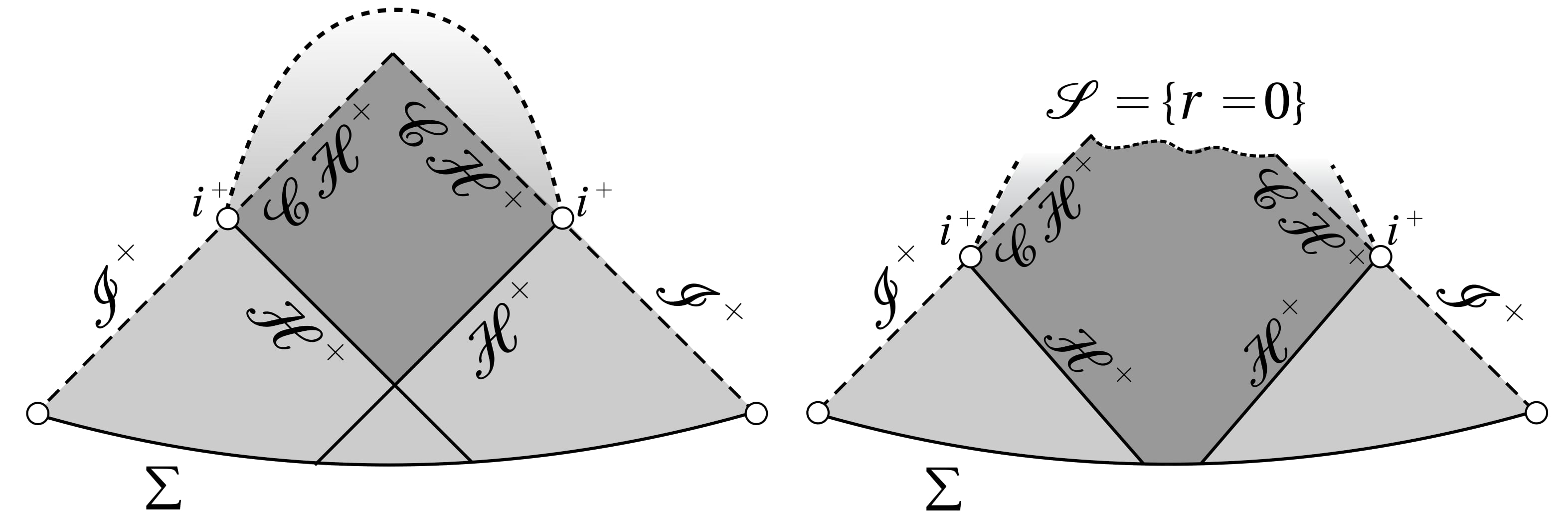}			\caption{Left: two-ended black hole with a Cauchy horizon but no spacelike singularity.\\ Right: two-ended black hole with a spacelike singularity $\mathcal{S}$ and a Cauchy horizon $\mathcal{CH}^+$.\\ In both cases, the spacetime can be continuously-extended through  $\mathcal{CH}^+$.}		\label{fig:twoended}	\end{figure}

The results  in Section~\ref{subsection:CHstab} are stated in terms of local regions of spacetime as depicted in Figure~\ref{fig:interior}; however, Conjecture~\ref{SCC.conj} is stated in terms of global inextendibility. In Penrose's original formulation \cite{PenroseSCC},  Strong Cosmic Censorship  is in the context of gravitational collapse. As we will see in Section~\ref{section.collapse} when discussing the global structure of collapse, this involves a one-ended spacetime with a regular center $\Gamma$.  However, outside of the study of gravitational collapse, it is also possible to consider the global structure of two-ended spacetimes, which also corresponds to that of the Reissner--Nordstr\"{o}m  \eqref{RN} or Kerr \eqref{Kerr} spacetime (see Figure~\ref{fig:kerr}). The  first global spherically-symmetric result for \eqref{EMSF} is due to Dafermos \cite{nospacelike} who proved that for \emph{small perturbations} of  Reissner--Nordstr\"{o}m metric, there is no spacelike singularity at all and only a null Cauchy horizon $\mathcal{CH}^+$ as depicted in the left panel of Figure~\ref{fig:twoended}. However, it follows  from   \cite{violent} that there exists  \emph{large perturbations} of the  Reissner--Nordstr\"{o}m metric obeying the assumptions of Theorem~\ref{CH.thm.SS} featuring  a non-empty spacelike singularity $\mathcal{S}=\{r=0\}$, as depicted on the right panel of Figure~\ref{fig:twoended}  (see Section~\ref{subsection.hairy}). For \eqref{Einstein} in vacuum, Dafermos--Luk \cite{KerrStab} considered  small perturbations of the Kerr metric and also showed the absence of spacelike singularities in this case. These results are summarized in the following theorem.
	
	\begin{Theorem}\label{SCC.thm} (Two-ended dynamical black holes with no spacelike singularity).
		~\begin{itemize}
			\item \cite{nospacelike,breakdown} Consider a bifurcate two-ended event horizon $\mathcal{H}^+$ on which the decay assumptions of Theorem~\ref{CH.thm.SS} are satisfied, and assume additionally that (the characteristic data induced by)  $(g,\phi)$ on $\mathcal{H}^+$ is a small perturbation of $(g_{RN},0)$ in an appropriate norm. Then the solution to \eqref{EMSF} (or \eqref{EMCSF}) has no spacelike singularity and its terminal boundary only consists of the (bifurcate) Cauchy horizon $\mathcal{CH}^+$,  as depicted  in the left  of Figure~\ref{fig:twoended}.
			\item \cite{KerrStab}  Consider a bifurcate two-ended event horizon $\mathcal{H}^+$ on which the decay assumptions of Theorem~\ref{CH.thm.vacuum}  are satisfied, and assume additionally that (the characteristic data induced by)   $g$ on $\mathcal{H}^+$ is a small perturbation of $g_{K}$ in an appropriate norm. Then the solution to \eqref{Einstein} in a vacuum has no spacelike singularity and its terminal boundary only consists of the  (bifurcate)  Cauchy horizon $\mathcal{CH}^+$,  as depicted  in the left  of Figure~\ref{fig:twoended}.
		\end{itemize}
	\end{Theorem} Returning to the issue of Strong Cosmic Censorship, we finally state the only known definitive statement of inextendibility for generic spherically symmetric two-ended solutions of \eqref{EMSF}.  \begin{Theorem}[$C^2$ Strong Cosmic Censorship in spherical symmetry \cite{JonathanStab,JonathanStabExt}] There exists an open and dense subset (for the topology induced by a smooth norm) within the set of  admissible spherically-symmetric initial
		data  such that their MGHD for \eqref{EMSF} is $C^2$ (future) inextendible.
		
	\end{Theorem} The above result by Luk--Oh  \cite{JonathanStab,JonathanStabExt}  can be interpreted as a full resolution of a spherically-symmetric version of Conjecture~\ref{SCC.conj} at the $C^2$ level (Statement~\ref{C2} of Section~\ref{subsection.MGHD}) for two-ended initial data. We emphasize that their $C^2$ extendibility result applies to both cases in Figure~\ref{fig:twoended}. Furthemore, we  note that a recent result of Sbierski \cite{JanC1} showing that the spacetimes of Theorem~\ref{SCC.thm} are also $C^1$ inextendible (which is an intermediate version of Strong Cosmic Censorship between Statement~\ref{H1} and Statement~\ref{C2} in Section~\ref{subsection.MGHD}). Finally, the Hawking mass blow-up (conditionally) obtained in Theorem~\ref{CH.instab.thm.SS} leads to conjecture the validity of a $H^1$ version Strong Cosmic Censorship \cite{Dafermos:2004jp}, i.e., Statement~\ref{H1} from Section~\ref{subsection.MGHD},  still open at this time.
	
	\subsection{Two-ended black holes with a spacelike singularity and a Cauchy horizon}\label{subsection.hairy}
	
	\hskip 3 mm \textbf{Interior of charged hairy black holes}. The author's works \cite{violent} and \cite{fluctuating} with Li considered the interior of   black holes with a scalar hair-- namely a non-trivial stationary  scalar field on the event horizon. These spacetimes are spherically-symmetric solutions of the Einstein--Maxwell--Klein--Gordon equations \eqref{EMKG} in \cite{violent} and Einstein--Maxwell-charged-scalar-field  \eqref{EMCSF} in \cite{fluctuating} and their terminal boundary is entirely-spacelike, thus they do not possess any Cauchy horizon (compare with Theorem~\ref{CH.thm.SS}). They have a two-ended  $\bbR \times \mathbb{S}^2$ spatial topology, but are not associated to an asymptotically flat\footnote{However, Zheng recently \cite{Weihao2} constructed a  class of asymptotically AdS (the analogue of asymptotic flatness in the presence of  a negative cosmological constant) black holes to which Theorem~\ref{hairy.thm} applies, see the discussion in Section~\ref{section.variant}.} black hole exterior, and thus  do not play any direct role in Strong Cosmic Censorship as formulated in Conjecture~\ref{SCC.conj}.  They will be used, however, in the asymptotically flat black hole constructions discussed   below. For now, we state the results of  \cite{violent,fluctuating}.
	\begin{Theorem}[\cite{violent,fluctuating}]\label{hairy.thm} There exists spatially-homogeneous solutions of the Einstein--Maxwell equations  coupled with a charged and/or massive scalar field, which is initially constant  $\phi \equiv \epsilon \neq0$ on a two-ended event horizon $\mathcal{H}_L \cup \mathcal{H}_R$ and obeying the following properties.
		
		\begin{enumerate}
			\item (violent collapse, \cite{violent}). For the Einstein--Maxwell-Klein--Gordon \eqref{EMKG} equations,  there exists a discrete set  $\mathcal{C}$ with $0\in \mathcal{C}$, such that, if $m^2 \in \mathbb{R}-\mathcal{C}$,  the MGHD terminates  at a spacelike  boundary $\mathcal{S}=\{r=0\}$ and $\mathcal{S}=\{r=0\}$  is approximated by a Kasner metric of exponents $(1-2 C \epsilon^2, C \epsilon^2, C \epsilon^2)$ for some $C>0$.
			
			\item\label{fluc} (fluctuating collapse, \cite{fluctuating}). For the Einstein--Maxwell-charged-scalar-field equations  \eqref{EMCSF}, if $q_0\neq 0$,  then the MGHD terminates  at a spacelike  boundary $\mathcal{S}=\{r=0\}$ and $\mathcal{S}=\{r=0\}$ is approximated by a Kasner metric of positive exponents $(1-2p(\epsilon),p(\epsilon),p(\epsilon))$, where $p(\epsilon) \in (0,\frac{1}{2})$ is an algebraic function  of  a highly-oscillating quantity  $\alpha(\epsilon)$ of the form $$\alpha(\epsilon) \approx  C \sin(\omega_0 \epsilon^{-2} +O(\log(\epsilon^{-1}))) \text{ as } \epsilon \rightarrow 0$$ for  $C\neq 0$ and $\omega_0 \in \mathbb{R}-\{0\}$. Moreover, for some values of $\epsilon$, a Kasner bounce takes place.
	\end{enumerate}	\end{Theorem}
	To clarify the terminology, we recall the Kasner spacetime \cite{Kasner}, a solution of the Einstein vacuum equations. For $(p_1,p_2,p_3) \in \mathbb{R}^3$ such that $p_1+p_2+p_3 =1$ (the Kasner exponents), define:  \begin{equation}\label{gK}
		g_K = -dt^2 + t^{2p_1} dx_1^2+t^{2p_2} dx_2^2+t^{2p_3} dx_3^2.
	\end{equation} If  $(p_1,p_2,p_3) = (1,0,0)$ (or perturbations), $g_K$ is flat (i.e., locally isometric to the Minkowski metric \eqref{minkowski}) but otherwise, $\mathcal{S}=\{t=0\}$ is a spacelike singularity. In Theorem~\ref{hairy.thm}, the Kasner exponents $p_i$ are all positive (which is consistent with the known stability results in this case \cite{FournodavlosRodnianskiSpeck}, see Section~\ref{section.variant}). Note, however, that in the uncharged-scalar-field case of \cite{violent}, the Kasner exponents  degenerate to the Minkowskian endpoint $(p_1,p_2,p_3)=(1,0,0)$ when $\epsilon$ becomes arbitrarily small. 
	
	The terminology ``Kasner bounce'' originates from the foundational  works of BKL \cite{BKL1,BKL2} on Kasner dynamics for the Einstein equations. These celebrated  heuristics proposed an instability mechanism for  Kasner metrics \eqref{gK} with one negative $p_i$ in three space-dimensions, leading to the transition (``Kasner epoch'') to a stable Kasner metric with all positive $p_i$'s.  Statement~\ref{fluc} of Theorem~\ref{hairy.thm} provides the first rigorous quantitative study  of the long-hypothesized
	Kasner bounce phenomenon; see also  the recent works \cite{Warreninv,Warreninv2} for the first examples of such Kasner bounces without assuming spatial homogeneity.
	
	\textbf{Construction of asymptotically flat two-ended black holes}. 
	A gluing argument allows to construct two-ended asymptotically flat black holes satisfying the assumptions of Theorem~\ref{CH.thm.SS} and thus with a non-empty Cauchy horizon $\mathcal{CH}^+$, but whose terminal boundary \emph{also} features a spacelike component $\mathcal{S}'$  isometric to a subset of  $\mathcal{S}=\{r=0\}$ in Theorem~\ref{hairy.thm} (see \cite{fluctuating}[Section 1.2] for details). In particular, one can in principle construct spherically-symmetric asymptotically flat solutions of Einstein--Maxwell-charged-scalar-field equations \eqref{EMCSF} exhibiting the Kasner bounce  described in the previous paragraph.	However, one would  additionally require decay estimates in the black hole exterior--which are open at present--to finalize the construction of two-ended asymptotically flat black holes featuring a null Cauchy horizon and a spacelike singularity; this is true  both in the massive case of  \cite{violent} (\eqref{EMKG})  and the charged case of \cite{fluctuating} (\eqref{EMCSF}), see the related discussions in Section~\ref{section.exterior}. 
	
	Independently of the constructions of Theorem~\ref{hairy.thm}, however, the author proved \cite{bif2} that there exist two-ended asymptotically flat black holes solutions of  Einstein--Maxwell-(uncharged)-scalar-field equations \eqref{EMSF} as in the right panel of  Figure~\ref{fig:twoended}, complementing Theorem~\ref{CH.instab.thm.SS} in showing both possibilities of Figure~\ref{fig:twoended} are possible.
	
	\begin{Theorem}[\cite{bif2}]\label{thm.2ended.construction}
	There exists a large class of spherically symmetric two-ended asymptotically flat black hole solution of  \eqref{EMSF}  such that the terminal boundary  has two non-empty components:    a weakly singular null Cauchy horizon $\mathcal{CH}^+  \neq~ \emptyset$, and  a crushing singularity $\mathcal{S}=\{r=0\}$. 		  Moreover,	$\mathcal{S}$ is spacelike near $\mathcal{CH}^+\cap \mathcal{S}$.

		\end{Theorem}

	We have not yet addressed the spacetime behavior  near the null-to-spacelike transition at $\mathcal{CH}^{+} \cap \mathcal{S}$ in  Theorem~\ref{thm.2ended.construction}. However, this transition is precisely characterized by the results of \cite{bif}, which are presented in the following Section~\ref{section.collapse} in the context of gravitational collapse, see Theorem~\ref{bif.thm}.

	\section{Gravitational collapse and (locally) naked singularities}\label{section.collapse}
	\subsection{Mathematical setting of gravitational collapse}\label{subsection.collapsedef}			
	Gravitational collapse is the widely-accepted theory accounting for the existence of astrophysical black holes, as pioneered in the celebrated work of Oppenheimer--Snyder \cite{OppenheimerSnyder} (recall the discussion in Section~\ref{subsection.open}). In this astrophysical process, a star whose initial state is free of any trapped surface develops a trapped surface at a later time. The mathematical setting for gravitational collapse imposes\footnote{Topological censorship considerations regarding an initial hypersurface $\Sigma$ indeed indicate that $\Sigma$ must have a trivial topology (like $\bbR^3$ in Definition~\ref{def.grav}) if it is free of any trapped surface, see  \cite{TC} for interesting results in this direction.} an initial surface $\Sigma$ that is diffeomorphic to $\bbR^3$ and free of any trapped surface, with the origin of  $\bbR^3$ corresponding to the center of  the star $\Gamma$. Of course, not all such spacetimes undergo a gravitational collapse process  
	(e.g. recall from Section~\ref{subsection.mink} that perturbations of the Minkowski spacetime do not admit any trapped surface). Nonetheless, in  Definition~\ref{def.grav} below,  we  introduce spacetimes that are initially free of any trapped surface: thus gravitational collapse may take place if the conditions leading to the formation of a trapped surface\footnote{There is a rich mathematical theory for the formation of trapped surface, see e.g.\ \cite{ChristoSCC,Christo1,trappedKLR,trappedKR,trappedAL,trappedeuler}.} are satisfied at a later time.	 For simplicity, we restrict ourselves to the vacuum case (\eqref{Einstein} with $\mathbb{T}=0$), keeping in mind that  Definition~\ref{def.grav}  generalizes in the presence of a suitable matter model (see e.g.\ \cite{Kommemi}).
	\begin{Def} \label{def.grav}We say a Lorentzian manifold $(\mathcal{M},g)$ solution of \eqref{Einstein} is a spacetime suitable for gravitational collapse  if it is the MGHD of suitable initial data set $(\Sigma,g_0,K_0)$ such that \begin{itemize}
			\item $\Sigma$ is diffeomorphic to $\bbR^3$ and free of any trapped surface. 
			\item $(g_0,K_0)$ are asymptotically flat, i.e., tends to the Minkowski solution $(m,0)$ 
			as $r\rightarrow+\infty$.
		\end{itemize}			The central axis $\Gamma$ is defined  as the wordline of the timelike geodesic starting  at the (diffeomorphic image of the) origin in  $\bbR^3$ in the future normal direction to $\Sigma$.
	\end{Def} Note that a spacetime suitable for gravitational collapse possesses one asymptotically flat end (``one-ended spacetime''), as opposed to the two-ended spacetimes (including the Reissner--Nordstr\"{o}m \eqref{RN} and Kerr \eqref{Kerr} MGHDs) discussed in Section~\ref{subsection.two} whose topology is that of $\bbR \times \mathbb{S}^2$.
	\subsection{From Weak Cosmic Censorship to Strong Cosmic Censorship}\label{subsection.christodoulou}
	In the celebrated series \cite{Christo1,Christo2,Christo3,ChristoCQG,Christo.existence}, Christodoulou studied a model of spherical gravitational collapse in considering spherically symmetric  solutions of  \eqref{ESF} satisfying the conditions of Definition~\ref{def.grav}. As we discussed  in Section~\ref{subsection.model}, such spacetimes cannot feature a Cauchy horizon emanating from $i^+$, but Christodoulou constructed \cite{Christo.existence} examples of (locally) naked singularity solutions with a Cauchy horizon $\mathcal{CH}_{\Gamma}$ emanating from the center $\Gamma$, as depicted in Figure~\ref{fig:naked}, with  initial data in a H\"{o}lder class.  Christodolou specifically studied a rougher class of spherically symmetric  solutions of  \eqref{ESF} within the BV (bounded variation) class in the series \cite{Christo1,Christo2,Christo3,Christo.existence}. He proved that  (locally) naked singularities satisfying the condition $\frac{\rho}{r} \not\to 0$ (recall the definition in \eqref{rho.def})  on their past light cone   are non-generic, in the sense they have finite co-dimension within this class. Combining the results of \cite{Christo1,Christo2,Christo3,Christo.existence}  with a sharp extension criterion given in \cite{ChristoCQG,Christo3} that implies  any naked singularity satisfies the condition $\frac{\rho}{r} \not\to 0$, results in the following theorem.
	
	\begin{Theorem}[\cite{Christo1,Christo2,Christo3,ChristoCQG,Christo.existence}]\label{WCC.theorem}		For spherically-symmetric initial data belonging to the BV class, the MGHD solution of \eqref{ESF} satisfy one the following:\begin{enumerate}[a.]
			\item\label{a} are geodesically complete, with the same  Penrose diagram as the Minkowski spacetime.
			\item\label{b} or possess a black hole region with a spacelike terminal boundary $\mathcal{S}=\{r=0\}$ with the Penrose diagram of Figure~\ref{fig:collapse}. In this case, the MGHD is $C^2$ (future) inextendible.
			\item or feature a (locally) naked singularity $b_{\Gamma}$ with the Penrose diagram of Figure~\ref{fig:naked}.
		\end{enumerate}
		Moreover, for generic initial data within the above class, only options \ref{a} or \ref{b} are possible.
	\end{Theorem}
	As is clear from Theorem~\ref{WCC.theorem}, the only obstruction to determinism for the above model are the Cauchy horizons $\mathcal{CH}_{\Gamma}$ emanating from the center $\Gamma$  which are, in principle, extendible. In view of Theorem~\ref{WCC.theorem} showing that spacetimes with $\mathcal{CH}_{\Gamma}\neq \emptyset$ are non-generic within the (spherically-symmetric) BV class, one can also interpret Theorem~\ref{WCC.theorem} as validating a version of Strong Cosmic Censorship at the $C^2$ level (Statement~\ref{C2} from Section~\ref{subsection.MGHD}) for rough and spherically-symmetric initial data.  We emphasize that, as in Section~\ref{sectio.blue}, Cauchy horizons constitute the main obstruction to Strong Cosmic Censorship. However, contrary to the Cauchy horizon $\mathcal{CH}^+$ emanating of $i^+$ discussed in  Section~\ref{sectio.blue} that is subjected to a blue-shift instability, there is no such mechanism  for Cauchy horizons $\mathcal{CH}_{\Gamma}$ emanating of $\Gamma$ and therefore, no hope for $\mathcal{CH}_{\Gamma}$  to be singular. That is why, should Strong Cosmic Censorship be true, it must be that $\mathcal{CH}_{\Gamma}=\emptyset$ for generic solutions.\begin{rmk}The strategy employed by Christodoulou \cite{Christo2,Christo3} to obtain $\mathcal{CH}_{\Gamma}=\emptyset$ for generic solutions is to perturb an hypothetical solution with $\mathcal{CH}_{\Gamma}\neq\emptyset$ in order to obtain a sequence of trapped surfaces converging to $b_{\Gamma}$: it is then easy to see that  $\mathcal{CH}_{\Gamma}=\emptyset$ for the perturbed solution \cite{Kommemi}.	\end{rmk} 
	
	While Theorem~\ref{WCC.theorem} only provides $C^2$ inextendibility, the spacelike nature of the terminal MGHD boundary $\mathcal{S}=\{r=0\}$  (and Theorem~\ref{C0schwarzschild.thm}) leads to conjecturing that a generic spherically-symmetric solution is $C^0$ inextendible. This expectation is reinforced by the  fact that  $\mathcal{S}=\{r=0\}$ is asymptotically Schwarzschild towards $i^+$ as proven in \cite{DejanAn} (see also \cite{AnZhang,Warren1}).
	For the more sophisticated Einstein--Maxwell-charged-scalar-field model \eqref{EMCSF} which additionally allows to study Cauchy horizons emanating from $i^+$ (as we explained in Section~\ref{subsection.model}), the global picture of spherically-symmetric gravitational collapse is more subtle and  several aspects will be discussed in Section~\ref{subsection.breakdown} and Section~\ref{subsection.spacelike}. We emphasize, however, that the Christodoulou argument showing that  $\mathcal{CH}_{\Gamma}=\emptyset$ generically is  local near $\Gamma$, and thus, independent of the dynamics near $i^+$. We may thus  speculate \cite{Kommemi,breakdown} that $\mathcal{CH}_{\Gamma}=\emptyset$  for generic spherically-symmetric solutions of the Einstein--Maxwell-charged-scalar-field model \eqref{EMCSF}.\subsection{The breakdown of  weak null singularities}\label{subsection.breakdown}
	
\begin{figure}[H]	\begin{center}\includegraphics[width=0.35\linewidth]{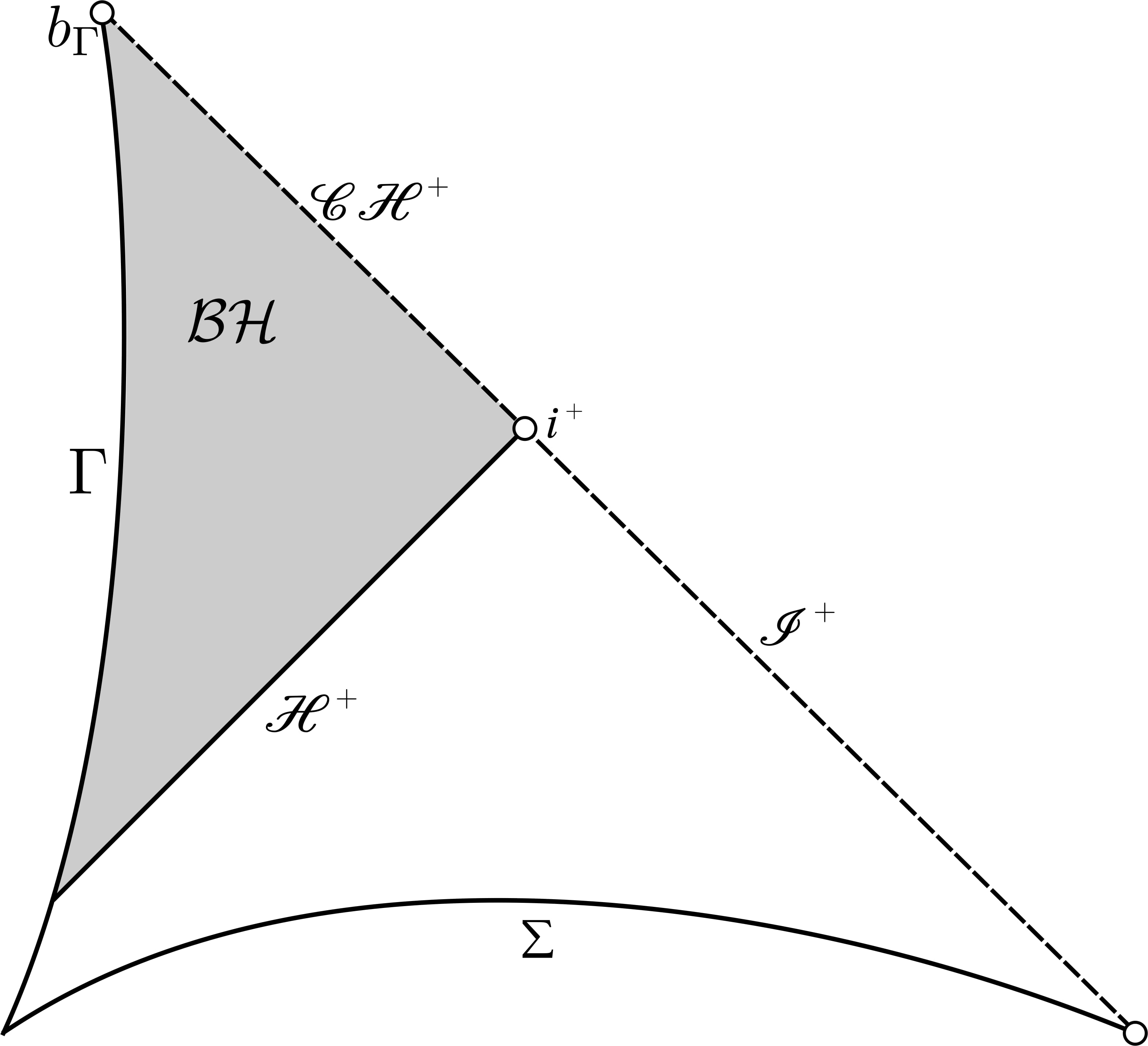}			\caption{Impossible  Penrose diagram 
				if $\mathcal{CH}^+$ is weakly singular by Theorem~\ref{breakdown.thm}.}		\label{fig:disprove}	\end{center}\end{figure}

	As we saw in Section~\ref{sectio.blue} (Theorem~\ref{CH.thm.SS}), the MGHD terminal boundary is not \emph{everywhere-spacelike} for spherically-symmetric solutions of the Einstein--Maxwell-charged-scalar-field model \eqref{EMCSF}  (at least, under decay conditions of the form \eqref{decay} on the event horizon that are conjectured to be satisfied for regular localized solutions). The falsification of Conjecture~\ref{spacelike.conj} for this model ironically begs the question: \emph{does there exist a spacelike singularity at all}? The two-ended spacetimes of Section~\ref{subsection.two}  provide examples in which $\mathcal{S}=\emptyset$ and the Cauchy horizon $\mathcal{CH}^+$ ``closes-off'' the spacetime, as depicted in  Figure~\ref{fig:twoended}, but, as explained in Section~\ref{subsection.collapsedef}, such spacetimes are not models of gravitational collapse and do not satisfy the assumptions of Definition~\ref{def.grav}. For one-ended spacetime suitable for gravitational collapse, it is possible, in principle, that  $\mathcal{CH}^+$ the Cauchy horizon emanating from $i^+$  is the only component of the MGHD terminal boundary and thus meets the central axis $\Gamma$ at its endpoint $b_{\Gamma}$ as depicted in Figure~\ref{fig:disprove}. In fact,  an example of a spacetime with the Penrose diagram of Figure~\ref{fig:disprove}  has been constructed in \cite{KehleUnger} using a gluing procedure. However, the author  proved that, providing the Cauchy horizon $\mathcal{CH}^+$ is weakly singular, it cannot ``close-off'' the spacetime, and therefore, the Penrose diagram of Figure~\ref{fig:disprove} is impossible \cite{breakdown} under the (conjecturally) generic conditions of gravitational collapse.
	\begin{Theorem}[Breakdown of weak null singularities \cite{breakdown}]\label{breakdown.thm}
		Let a spacetime suitable for gravitational collapse in the sense of Definition~\ref{def.grav} with a future terminal boundary $\mathcal{B}$ including  a Cauchy horizon $\mathcal{CH}^+\subset\mathcal{B}$ emanating from $i^+$. Assume  that $ \mathcal{CH}^+$ is weakly singular, in the sense of Theorem~\ref{CH.instab.thm.SS}. Then $\mathcal{CH}^+\underset{\neq}{\subset}\mathcal{B}$, i.e., the Penrose diagram of the spacetime is not given by Figure~\ref{fig:disprove}.
	\end{Theorem} 
	It is important to point out that the examples of \cite{KehleUnger} with the Penrose diagram of Figure~\ref{fig:disprove} have a non-singular Cauchy horizon $\mathcal{CH}^+$ (consistently with Theorem~\ref{breakdown.thm}), a situation which is conjectured to be non-generic. Indeed, in Theorem~\ref{CH.instab.thm.SS}, we recall that $\mathcal{CH}^+$ was shown to be weakly singular under decay assumptions on the  event horizon that are conjectured to be generic. The main mechanism in Theorem~\ref{breakdown.thm} -- the  ``breakdown of the weak null singularity'' $\mathcal{CH}^+$ -- is proved  by a contradiction argument, using the focusing properties of the Einstein equations (notably the null constraint, so-called Raychaudhuri, equations) and taking advantage of the strength of the singularity at $\mathcal{CH}^+$ combined with the spacetime geometry near the central axis $\Gamma$.  In view of the breakdown of $\mathcal{CH}^+$, there must be another (presumably spacelike, see Figure~\ref{fig:spacelikeconj}) component of the terminal boundary; we  will discuss this aspect
	in Section~\ref{subsection.spacelike} below. 
	
	\subsection{The spacelike portion of the terminal boundary}\label{subsection.spacelike}
\begin{figure}[H]	\begin{center}\includegraphics[width=0.4\linewidth]{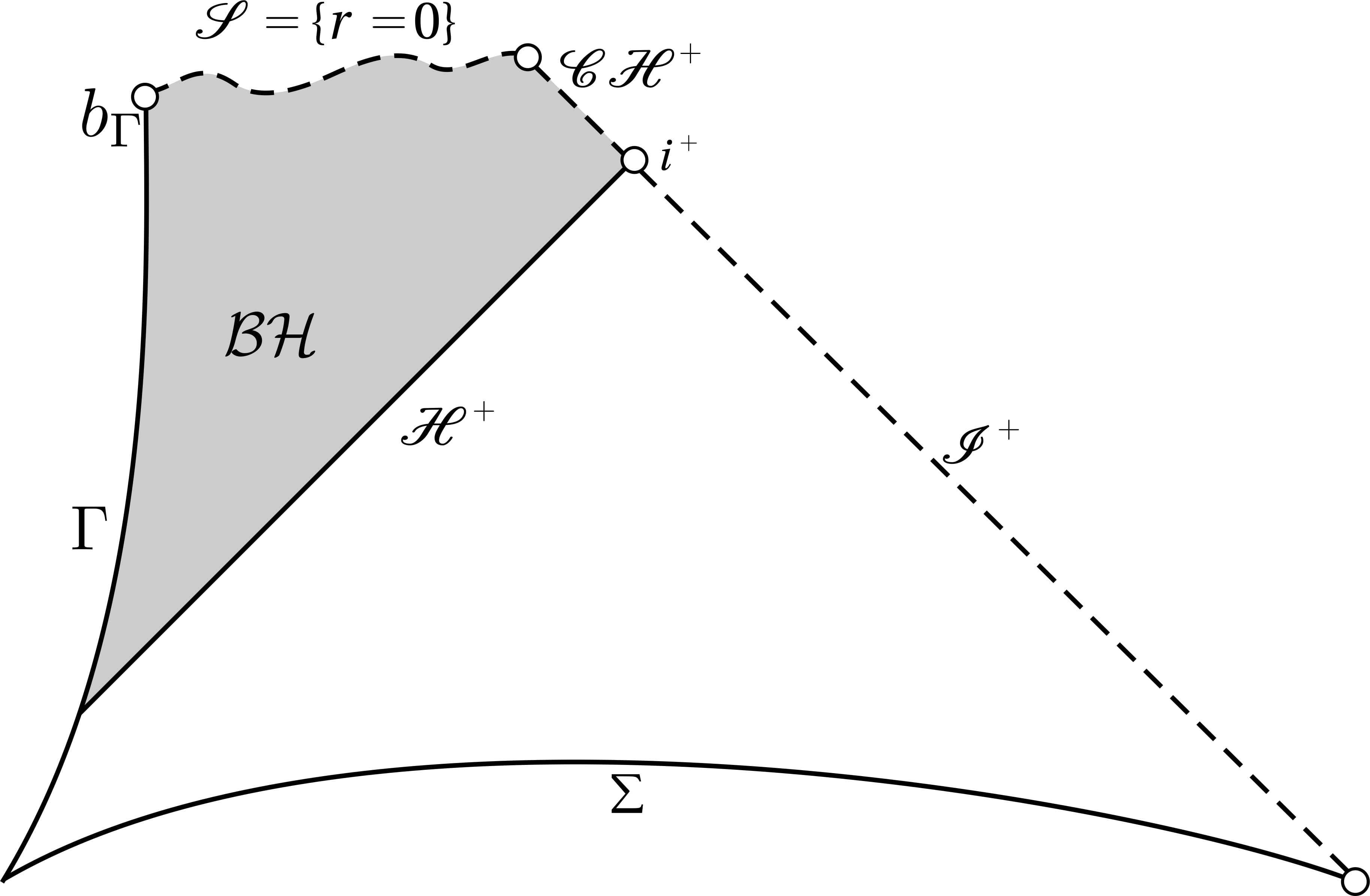}			\caption{\small Penrose diagram of the one-ended black holes of 
			Theorem~\ref{bif.thm2}, with $\mathcal{S}=\{r=0\}$. \normalsize}		\label{fig:spacelikeconj}	
			\end{center}\end{figure}

	 Theorem~\ref{breakdown.thm} is consistent with a spacetime geometry whose Penrose diagram is the rightmost in Figure~\ref{fig:naked}, i.e., with a Cauchy horizon $\mathcal{CH}_{\Gamma}$ emanating from a locally naked singularity $b_{\Gamma}$. Barring this possibility, however, Theorem~\ref{breakdown.thm} shows that $\mathcal{B}-\mathcal{CH}^+= \mathcal{S}=\{r=0\}$, where $\mathcal{S}$ is not entirely null and presumably spacelike \cite{breakdown}.  The conjectured non-genericity of locally naked singularities discussed in Section~\ref{subsection.christodoulou} together with Theorem~\ref{breakdown.thm}  then imply that a generic spherically-symmetric black hole solution of the Einstein--Maxwell-charged-scalar-field system \eqref{EMCSF} has the Penrose diagram of Figure~\ref{fig:spacelikeconj}, as was originally conjectured by Dafermos \cite{Dafermos:2004jp}. Note the presence  of two coexisting singular components of the terminal boundary of different strengths, namely $\mathcal{CH}^+$, which is a weak singularity (blow-up of curvature) and  $\mathcal{S}=\{r=0\}$, which is a stronger singularity. The quantitative study of the transition region between $\mathcal{CH}^+$  and $\mathcal{S}$ in Figure~\ref{fig:spacelikeconj} was carried out in the author's recent work \cite{bif}, under the assumption that the scalar field does not  oscillate\footnote{As it turns out, this assumption is always satisfied in the uncharged scalar field model \eqref{EMSF}, see Section~\ref{section.exterior}.} too quickly towards the Cauchy horizon $\mathcal{CH}^+$.  The main result is that $\mathcal{S}=\{r=0\}$ is indeed spacelike\footnote{In particular, there is no null ingoing boundary component $\mathcal{S}_{i^+}$ on which $r=0$ that extends the Cauchy horizon $\mathcal{CH}^+$, see \cite{bif}.} in the vicinity of $\mathcal{CH}^+$  and described by a Kasner-type metric with variable Kasner exponents (see \cite{FournodavlosLuk,bif} for a definition) that degenerate towards the intersection  $\mathcal{CH}^+\cap \mathcal{S}\subset \{v=\infty\}$, where $v$ is the same advanced-time  coordinate as in \eqref{decay}.
	\begin{Theorem}[\cite{bif}]\label{bif.thm}
		Assume the spacetime  in Theorem~\ref{breakdown.thm} does not possess a locally naked singularity, i.e., $\mathcal{CH}_{\Gamma}=\emptyset$, and  denote  $\mathcal{B}$ its terminal boundary.		Then $\mathcal{B}$ contains a  spacelike singularity $\mathcal{S} \neq \emptyset$ intersecting $\mathcal{CH}^+\subset\{v=\infty\}$   and the metric near $\mathcal{CH}^+ \cap \mathcal{S}$ is approximated by a Kasner metric of $v$-dependent Kasner exponents $(1-2p(u,v),p(u,v),p(u,v))$, where $p(u,v)\approx \frac{C}{v}$ as $v\rightarrow \infty$ for some $C>0$.
	\end{Theorem}
While Theorem~\ref{bif.thm} in a local result in the black hole interior,  the companion paper \cite{bif2} provides concrete  global examples of one-ended asymptotically flat spacetimes satisfying the assumptions of Theorem~\ref{bif.thm}: 	\begin{Theorem}[\cite{bif2}] \label{bif.thm2}
		There exist examples of spherically symmetric one-ended asymptotically flat black hole solution of  \eqref{EMCSF}  such that the terminal boundary  has two non-empty components:    a weakly singular null Cauchy horizon $\mathcal{CH}^+  \neq~ \emptyset$, and  a crushing singularity $\mathcal{S}=\{r=0\}$. 		  Moreover,	$\mathcal{S}$ is spacelike near $\mathcal{CH}^+\cap \mathcal{S}$ and obeys the Kasner estimates of Theorem~\ref{bif.thm}.

	\end{Theorem} 
	\begin{rmk} Strictly speaking, the analysis of Theorem~\ref{bif.thm} is unnecessary to prove a spherically-symmetric version of Strong Cosmic Censorship for the charged  collapse model, since  \cite{Kommemi}  the only possible boundary components in spherical symmetry are the Cauchy horizon $\mathcal{CH}^+$ and boundary components on which $r=0$  (excluding the conjecturally non-generic locally naked singularities $\mathcal{CH}_{\Gamma}$).  $C^2$ extensions through  $\mathcal{CH}^+$ are impossible  by Theorem~\ref{CH.instab.thm.SS}, and through components  on which $r=0$,   due to the blow-up of the Kretschmann scalar \cite{Kommemi}.				 				  However, a complete resolution of Strong Cosmic Censorship  \emph{without spherical symmetry}  requires a comprehensive quantitative understanding of the coexistence of weak and strong singularities  depicted in Figure~\ref{fig:spacelikeconj}.
		
	\end{rmk} Theorem~\ref{bif.thm} provides the first examples of a null-spacelike singularity transition\footnote{Theorem~\ref{bif.thm} is, in fact, the corollary of a more general local result in \cite{bif}. In particular, the Kasner-like quantitative estimates of Theorem~\ref{bif.thm}  also apply to the two-ended black holes with $\mathcal{S}\neq \emptyset$ as in the rightmost diagram of Figure~\ref{fig:twoended}, see \cite{bif2}.} in a black hole, together with the first construction a Kasner-like singularity with degenerating exponents (see \cite{FournodavlosLuk} for a very general  construction of Kasner-type  spacetimes, which however assumes the Kasner exponents to be non-degenerating and thus does not cover the situation of Theorem~\ref{bif.thm}). Lastly, one may speculate that the results of Theorem~\ref{bif.thm} can be suitably generalized\footnote{The landmark result of Fournodavlos--Rodnianski--Speck \cite{FournodavlosRodnianskiSpeck}  indeed proves that Kasner metrics with constant positive exponents are stable to small perturbations. The Kasner exponents in Theorem~\ref{bif.thm} are also positive,  thus the spacetime is  expected  to be stable against non-spherical perturbations. However, the Kasner exponents in Theorem~\ref{bif.thm} are non-constant and degenerate as $v\rightarrow +\infty$, and therefore the techniques  of \cite{FournodavlosRodnianskiSpeck} do not apply.}  to non spherically-symmetric spacetimes solving the Einstein-scalar-field system \eqref{Einstein}, \eqref{ESF}. 
	In summary, it is conjectured that the  geometry of generic (non spherically-symmetric) Einstein-scalar-field solutions is as in Figure~\ref{fig:spacelikeconj}. The situation for the  Einstein vacuum equations, however, is more uncertain in that Kasner-type singularities  are not expected to be stable, see \cite{FournodavlosLuk,ringstrombianchi}.
	
	\section{Late-time tails in the black hole exterior and connections to black hole stability}\label{section.exterior}
	In this section, we discuss decay results on the event horizon $\mathcal{H}^+$ of the form \eqref{decay}. We have  emphasized in Section~\ref{sectio.blue} that  decay assumptions are essential in the black interior dynamics (recalling Theorems~\ref{CH.thm.SS}, \ref{CH.osc.thm}, \ref{CH.thm.vacuum} and \ref{CH.instab.thm.SS}). In particular, upper bounds are used in stability results of the Cauchy horizon $\mathcal{CH}^+$ and lower bounds to show its (weakly) singular character. Proving Strong Cosmic Censorship starting from generic initial data in the black hole exterior, as prescribed in Conjecture~\ref{SCC.conj},  requires proving that such upper and lower bounds are satisfied on the event horizon $\mathcal{H}^+$  for generic (smooth) initial data. We  discuss  results in this direction below, starting from the toy problem provided by  linear wave-type equations on a fixed black hole.				\subsection{Linear wave equations on a stationary black hole and Price's law}\label{subsection.linear}
	Inverse-polynomial time decay for the wave equation \begin{equation}\label{wave}
		\Box_g \phi=0
	\end{equation} on a black hole spacetime has been conjectured since the numerics of	Price \cite{Pricepaper} articulating what is now known as Price's law. Regarding mathematical works, boundedness and integrated decay  for \eqref{wave} on a Schwarzschild  $g=g_S$ or (sub-extremal) Reissner--Nordstr\"{o}m  $g=g_{RN}$ black hole  were established in \cite{BlueSoffer1,Red} and  \cite{Blue,Red,claylecturenotes,Tataru,KerrDaf,hidden} on the (sub-extremal) Kerr black hole $g=g_{K}$.  Decay upper bounds were proven   on Schwarzschild/Reissner--Nordstr\"{o}m \cite{Schlag2,Schlag1} and  Kerr \cite{Tataru,Tataru2}. Precise asymptotic tails were subsequently obtained by Hintz \cite{Hintz} and Angelopoulos--Aretakis--Gajic  \cite{AAG1,AAG2}. While these tails are valid throughout the black hole exterior, we will only state them on the event horizon $\mathcal{H}^+$ in the theorem below for simplicity, using an Eddington--Finkelstein advanced-time coordinate $v$ on  $\mathcal{H}^+$ as in Section~\ref{sectio.blue}.
	\begin{Theorem}[Price's law, \cite{AAG1,Hintz,AAG2}]\label{Pricelaw.thm} Let $g$ be either a  Schwarzschild  $g=g_S$ or  (sub-extremal) Reissner--Nordstr\"{o}m black $g=g_{RN}$  or (sub-extremal) Kerr $g=g_{K}$ black hole exterior and $\phi$ a solution of \eqref{wave} with smooth compactly supported initial data at $\{t=0\}$. Then, there exists a function $I_0(\theta,\varphi)$  on $\mathbb{S}^2$, $\epsilon>0$,  such that $\phi$ obeys the following asymptotics as $v\rightarrow +\infty$: \begin{equation}\label{decay.precise}
			\phi_{|\mathcal{H}^+}(v,\theta,\varphi) = I_0(\theta,\varphi) v^{-3}  + O( v^{-3-\epsilon}).
		\end{equation} Moreover,  $I_0(\theta,\varphi) \not \equiv 0$ for generic initial data.
	\end{Theorem} The proof of \eqref{decay.precise} (and estimates for higher order derivatives)  in Theorem~\ref{Pricelaw.thm} show that the assumptions of Theorem~\ref{linear.int.thm} are satisfied for generic smooth and localized solutions of \eqref{wave}. 
	
	While		\eqref{wave} is often considered as a linear toy model for the Einstein equations \eqref{Einstein} in a vacuum,    the equations of linearized gravity 
	around the Kerr metric \eqref{Kerr} give rise to the celebrated Teukolsky equations			\cite{Teukolsky} which are spin-weighted analogues of \eqref{wave} and have the following form on (sub-extremal) Kerr spacetime \eqref{Kerr}, where the spin $s$ takes the value $\{\pm 2\}$. \begin{equation}\label{teukosly.eq}
		\begin{split}
			&\bigg[\Box_{g_K} +\frac{2s}{r^2+a^2\cos^2(\theta)}(r-M)\partial_r +\frac{2s}{r^2+a^2\cos^2(\theta)}(\frac{a(r-M)}{(r-r_+)(r-r_-)}+i\frac{\cos(\theta)}{\sin^2(\theta)})\partial_{\varphi}\\
			&+\frac{2s}{r^2+a^2\cos^2(\theta)}(\frac{M(r^2-a^2)}{(r-r_+)(r-r_-)}-r -ia\cos(\theta))\partial_t +\frac{1}{r^2+a^2\cos^2(\theta)}(s-s^2\cot^2(\theta))\bigg]\alpha_{s}  =0.
		\end{split}
	\end{equation} Boundedness and integrated decay for \eqref{teukosly.eq} have been established in  \cite{SRTdC2020boundedness,SRTdC2023boundedness2,TeukolskyDHR,MaKerr,AB15}, and  late-time tail results analogous to Theorem~\ref{Pricelaw.thm} were obtained in \cite{millet,Ma4,Ma5}; see  \cite{Ma1,pasqualotto2019spin,Maxwell3,TataruMaxwell,BlueMaxwell,Giorgi.RN} for analogous results on  Schwarzschild/Reissner--Nordstr\"{o}m black holes (namely,  the analogue of \eqref{teukosly.eq} with $a=0$).\color{black}
	
	\subsection{Charged scalar field equation} Motivated by the study of the Einstein--Maxwell-charged-scalar-field equations discussed in Section~\ref{sectio.blue} and Section~\ref{section.collapse}, we  introduce the charged scalar field equivalent of \eqref{wave} on a fixed black hole consisting of the following nonlinear Maxwell-charged-scalar-field system of equations: \begin{equation}\label{wave.CSF}
		g^{\mu \nu} D_{\mu } D_\nu \phi=0,\ \nabla^{\mu} F_{\mu \nu}=i q_0 \Im(\bar{\phi} D_{\nu} \phi),\  D_{\mu} = \nabla_{\mu}+ i q_0 A_{\mu},\ F= dA.
	\end{equation} For  spherically-symmetric  solutions of \eqref{wave.CSF} on a  black hole metric $g$ (like \eqref{schw} or  \eqref{RN}), we define the local charge $Q$ (a function on spacetime) and  the  asymptotic  charge $e\in \mathbb{R}$ as follows:  \begin{equation}\label{e.def}
		e = \lim_{v\rightarrow +\infty} Q_{|\mathcal{H}^+}(v), \text{ where } F= \frac{Q}{r^2} \Omega^2 du \wedge dv \text{ and } g= -\Omega^2 du dv + r^2 (d\theta^2 + \sin^2(\theta) d\varphi^2).
	\end{equation} One should expect $e\neq 0$ generically. As it turns out,  for $g=g_{S}$ or $g=g_{RN}$ (in the sub-extremal case),	$g^{\mu \nu} D_{\mu } D_\nu \phi=0$ asymptotically resembles a wave equation with an inverse-square potential \cite{Moi2,MoiDejan}, where the potential is of the schematic form $V(r) \approx iq_0 e\ r^{-2}$. Heuristics in the physics literature \cite{HodPiran1,HodPiran2} have led to  the following conjectured asymptotic tail for solutions of \eqref{wave.CSF}
	
	\begin{equation}  \label{conj.decay.2.eq}
		\phi_{|\mathcal{H}^+}(v,\theta,\varphi) = C_H \cdot e^{ \frac{iq_0 e}{r_+} v} \cdot v^{-1-\sqrt{1-4(q_0 e)^2}}+ O(v^{-1-\epsilon}),
	\end{equation}  where $C_H \neq 0$ for generic solutions and $\epsilon>0$. Note that the decay condition of the second part of Theorem~\ref{CH.thm.SS} is satisfied by \eqref{conj.decay.2.eq}, as is the oscillation condition \eqref{osc} required in Theorem~\ref{CH.osc.thm}. While the proof of \eqref{conj.decay.2.eq} is still open for spherically-symmetric solutions of \eqref{wave.CSF}, sharp upper bounds corresponding to \eqref{conj.decay.2.eq} have been established in the limit $|q_0 e|\ll 1$ by the author \cite{Moi2}. \begin{Theorem}[\cite{Moi2}]
		Spherically-symmetric solutions  of \eqref{wave.CSF} on a sub-extremal Reissner--Nordstr\"{o}m metric $g=g_{RN}$  with smooth, localized and small initial data at $\{t=0\}$ obey decay upper bounds of the form \eqref{decay} for some $s(q_0e)>1$, and $s(q_0e) \rightarrow 2$ as $q_0 e \rightarrow 0$.
	\end{Theorem} We also refer to \cite{MoiDejan} for the proof of late-time tails of the form \eqref{conj.decay.2.eq} for a toy-model of \eqref{wave.CSF}.
	\subsection{The Klein--Gordon equation for massive scalar fields}\label{subsection.KG} We now turn to the Klein--Gordon equation \eqref{KG} (with $m^2>0$), which is the massive variant of the wave equation \eqref{wave}  (noting that  Theorem~\ref{CH.thm.SS}, Theorem~\ref{CH.osc.thm} and Theorem~\ref{CH.instab.thm.SS}  still apply to this case). We  emphasize that the picture of Strong Cosmic Censorship for the Einstein--Klein--Gordon equations is conjectured to be qualitatively different from the vacuum case.  Some solutions of \eqref{KG} on a (sub-extremal) Kerr spacetime indeed  do not decay, but instead grow exponentially in time (growing mode solution), as proven by Shlapentokh-Rothman \cite{Yakov} (see also  \cite{Detweiler,Superradiance} for pioneering  works  in the Physics literature, together with  the review \cite{superadiance}).
	\begin{Theorem}[Kerr superradiant instability, \cite{Yakov}] Fix a  sub-extremal Kerr \eqref{Kerr} metric.  For an open set of $m^2>0$, there exist growing mode solutions of \eqref{KG} with smooth, localized initial data.
	\end{Theorem} In view of the instability of the Kerr metric, one would need to identify an (hypothetical) generic endstate\footnote{It has been debated \cite{SantosWCC,Superradiance,Detweiler,hair.unstable} whether  the Schwarzschild black hole  may  be a possible endstate. Note also that  the non-trivial stationary black holes solutions of the Einstein--Klein--Gordon equations (so-called ``hairy black holes'')  constructed by Chodosh--Shlapentokh-Rothman \cite{chodosh-sr1,chodosh-sr2} are also hypothetical endstate candidates. Lastly, it is also possible that \emph{some} sub-extremal Kerr black hole are immune to the superradiant instability, and thus could be stable \cite{Yakov}.} of Klein--Gordon collapse and show its inextendibility as a first step to proving Conjecture~\ref{SCC.conj}, a problem which is completely open at present. However, the Klein--Gordon equation \eqref{KG} on a Schwarzschild \eqref{schw} or (sub-extremal) Reissner--Nordstr\"{o}m  \eqref{RN} black hole, in contrast to the Kerr case, decays on the event horizon $\mathcal{H}^+$ at an inverse-polynomial rate of the form \eqref{decay} with $s=\frac{5}{6}$ as proven by Pasqualotto, Shlapentokh-Rothman and the author \cite{KGSchw1,KGSchw2}.
	\begin{Theorem}[\cite{KGSchw1,KGSchw2}]\label{KG.thm}
		Let $\phi$ be a solution of the Klein--Gordon equation \eqref{KG} with $m^2>0$  on a fixed   Schwarzschild \eqref{schw} or (sub-extremal) Reissner--Nordstr\"{o}m  \eqref{RN} black hole with smooth and compactly supported initial data at $\{t=0\}$. Then for every $\epsilon>0$, there exists $D>0$ such that: \begin{equation*}
			|\phi_{|\mathcal{H}^+}|(v,\theta,\varphi) \leq D \cdot v^{-\frac{5}{6}+\epsilon}.
		\end{equation*} Moreover, there exists $C \in \mathbb{R}$ such that, if $\phi$ has spherically-symmetric initial data, for every $\delta>0$ \begin{equation}\begin{split}\label{KG.decay}
				& \bigl| \phi_{|\mathcal{H}^+}(v,\theta,\varphi)- F(v) v^{-\frac{5}{6}} \bigr|   \leq D \cdot v^{-1+\delta},\\ & F(v) = C \sum_{q=1}^{+\infty} \gamma^{q-1} q^{\frac{1}{3}} \cos( \tilde{\Psi}_q(v)),\  \tilde{\Psi}_q(v) = m v   -\frac{3}{2} [2\pi M]^{\frac{2}{3}} m q^{\frac{2}{3}}  v^{\frac{1}{3}}  + \phi_-(M,e,m^2)+O(v^{-\frac{1}{3}}),
			\end{split}
		\end{equation} where $|\gamma|(M,e,m^2)<1$ and $\phi_-(M,e,m^2)\in \bbR$. Moreover, $C\neq 0$ for generic initial data. 
	\end{Theorem}
	In the spherically symmetric case, $\phi_{|\mathcal{H}^+}$ satisfies decay upper and lower bounds analogous to the ones required in Theorem~\ref{CH.thm.SS} and Theorem~\ref{CH.instab.thm.SS}; moreover, the leading order term  $F(v) v^{-\frac{5}{6}}$ satisfies the oscillation condition \eqref{osc} (with $\omega_{res}=0$, since $q_0=0$ in Theorem~\ref{KG.thm}) corresponding to the one required in  Theorem~\ref{CH.osc.thm}. In view of Theorem~\ref{KG.thm}, we conjecture that \eqref{KG.decay} also holds for spherically-symmetric   solutions for the Einstein--Klein--Gordon model with regular and localized initial data; proving such a statement is indeed key to a complete resolution of the spherically-symmetric version of Strong Cosmic Censorship for the (massive) Einstein--Klein--Gordon model.
	
	\subsection{Wave equations on dynamical black holes and black hole stability}
	An important pre-requisite to Strong Cosmic Censorship in the form of Conjecture~\ref{SCC.conj}  is the proof of asymptotic  stability of the black hole exterior region. In the simpler setting of spherical symmetry, this was achieved by Dafermos--Rodnianski \cite{PriceLaw} for the Einstein--Maxwell-scalar-field system \eqref{EMSF}, who also proved upper bounds (Price's law) analogous to \eqref{decay.precise} on the event horizon  that are sufficient to satisfy the assumptions of Theorem~\ref{CH.thm.SS}.  Luk--Oh \cite{JonathanStabExt} subsequently proved that   ($L^2$-averaged)
	inverse-polynomial lower bounds hold on the event horizon, thus showing that the assumptions of Theorem~\ref{CH.instab.thm.SS} are satisfied for spherically-symmetric solutions of \eqref{EMSF} with generic, smooth and localized  initial data at $\{t=0\}$.  Recently Luk--Oh \cite{twotails} and Gautam \cite{Gautam}   obtained asymptotic point-wise tails (hence point-wise lower bounds). We summarize these results below.
	\begin{Theorem}[\cite{PriceLaw,JonathanStabExt,twotails,Gautam}]\label{EMSF.ext.thm} Given any asymptotically flat,					spherically symmetric, admissible data  on $\mathbb{R}\times \mathbb{S}^2$, the MGHD $(\mathcal{M},g,\phi)$ for \eqref{EMSF} converges to a Reissner--Nordstr\"{o}m black hole at the following rate: there exists $D\in \mathbb{R}$, $\epsilon>0$, $C\in \mathbb{R}$ such that \begin{equation}
			\bigl|	\phi_{|\mathcal{H}^+}(v,\theta,\varphi)- D \cdot v^{-3}  \bigr| \leq C \cdot  v^{-3-\epsilon}.
		\end{equation} Moreover, $D\neq 0$ for generic initial data.
		
	\end{Theorem} \begin{rmk} Note that no smallness assumption  is imposed on the initial data in Theorem~\ref{EMSF.ext.thm}.  However, the admissibility condition requires the presence of a trapped surface at $\{t=0\}$, therefore the dynamical solution cannot disperse to the Minkowski spacetime.
	\end{rmk}
	Returning to the vacuum Einstein equations without symmetry, the proof of stability of the Kerr black hole exterior \eqref{Kerr} was obtained  in the case $|a|\ll M$ in the work of Giorgi, Klainerman, Shen and Szeftel  \cite{klainerman2021kerr,stabilitykerrformalism,GCM,effectiveuniform,ShenGCM,GKS}; see the closely related work of Dafermos--Holzegel--Rodnianski--Taylor \cite{SchwarzschildStab} on the stability of Schwarzschild, as well as \cite{blackbox,KS.polarized,SRTdC2020boundedness,SRTdC2023boundedness2,Giorgi.RN,Blue,TeukolskyDHR,HHV} for related linear and nonlinear black hole stability results. The  decay upper bounds obtained on the event horizon $\mathcal{H}^+$ of the dynamical Kerr perturbations satisfy the assumptions of Theorem~\ref{CH.thm.vacuum}: therefore, small perturbations of the  Kerr  exterior \eqref{Kerr} (for $|a|\ll M$) admit a $C^0$-extendible Cauchy horizon as depicted in Figure~\ref{fig:interior}. Linearized results  \cite{SRTdC2020boundedness,SRTdC2023boundedness2}  on solutions of \eqref{teukosly.eq}  indicate that the assumptions of Theorem~\ref{CH.thm.vacuum} are also  satisfied for small perturbations of the  Kerr  exterior \eqref{Kerr} in the full sub-extremal range $|a|< M$. We emphasize, however, that a full resolution of Conjecture~\ref{SCC.conj} requires to understand the generic endstate of gravitational collapse, which may involve several black holes. In that respect, we mention the celebrated \emph{final state conjecture} \cite{Penrose1982}, postulating that a generic solution of \eqref{Einstein} consists of finitely many Kerr black holes moving away from each others. Understanding spacetimes with multiple black-hole  is thus a necessary preliminary step prior to  a full proof of Strong Cosmic Censorship. 
	
	We have already emphasized that decay lower bounds on a black hole's event horizon are necessary to prove the (weakly) singular character of Cauchy horizon $\mathcal{CH}^+$: proving such lower bounds naturally requires to understand the late-time tail of wave equations on a dynamical black hole, an endeavor going beyond the problems of Section~\ref{subsection.linear}. For \eqref{wave}, it is known that  the Price's law asymptotics of Theorem~\ref{Pricelaw.thm} also hold on a class of dynamical black holes \cite{Tataru2,twotails}. However, it turns out that for other wave equations, the late-time tails are in general \emph{different on dynamical black holes} (at least in odd space-dimensions) as recently established by Luk--Oh \cite{twotails}. While their result provides  late-time asymptotics for a  general class of wave equations in odd space dimensions, for concreteness, we focus on the following example of \eqref{wave} on a spherically-symmetric dynamical black hole $\tilde{g}$ converging (sufficiently fast) to  the Schwarzschild exterior $g_S$ (in $(3+1)$ spacetime dimensions). Defining $\Pi_{\geq \ell}$ as the projection on  spherical harmonics of order $\ell$ and higher and assuming compactly supported initial data, the result of Luk--Oh \cite{twotails} provides the following dichotomy, under the assumption that the metric $\tilde{g}$ is non-stationary and $\ell\geq 1$:  \begin{itemize}
		\item  If $\Box_{\tilde{g}}\phi=0$, then $\Pi_{\geq \ell}\phi_{|\mathcal{H}^+}(v,\theta,\varphi) \sim C_{\ell}(\theta,\varphi) v^{-2 -2\ell} 
		$, and $C_{\ell} \neq 0$ for generic initial data.
		\item If $\Box_{g_S}\phi=0$, then $\Pi_{\geq \ell}\phi_{|\mathcal{H}^+}(v,\theta,\varphi) \sim C_{\ell}(\theta,\varphi) v^{-3 -2\ell} 
		$, and $C_{\ell} \neq 0$ for generic initial data.
	\end{itemize}
	The asymptotics of higher order spherical harmonics have no direct relevance to Strong Cosmic Censorship. However, there is a strong resemblance between by $\Pi_{\geq 2} \phi$, where $\phi$ solves \eqref{wave} and the Teukolsky equation \eqref{teukosly.eq} for $s=\pm 2$. Motivated by this analogy, Luk--Oh conjectured \cite{twotails} that generic black hole solutions of the vacuum Einstein equations   approaching a sub-extremal Kerr metric decay at the rate $v^{-6}$ on the event horizon $\mathcal{H}^+$, as opposed to the $v^{-7}$ rate proven for solutions of \eqref{teukosly.eq} on the exactly stationary Kerr exterior \cite{Ma5}. Proving such  asymptotic tails  on a dynamical black hole will be a crucial step towards a proof of Conjecture~\ref{SCC.conj}.

	\section{Strong Cosmic Censorship Conjecture beyond the context of gravitational collapse}\label{section.variant} We have focused so far on   aspects of Strong Cosmic Censorship in generic $(3+1)$-dimensional gravitational collapse (except in Section~\ref{subsection.two} and Section~\ref{subsection.hairy}), the context  in which it was originally formulated by Penrose \cite{PenroseSCC}. In this section, we discuss Strong Cosmic Censorship in other contexts, together with related questions that are in principle excluded by the genericity condition in Conjecture~\ref{SCC.conj}, but still relate to the dynamics of black holes  and their interior singularities.
	
	\subsection{Exceptional  dynamical black holes arising from  finite co-dimension initial data}
	
	We start remarking that the one-parameter  Schwarzschild family \eqref{schw} is a co-dimension three (respectively one) sub-family of the Kerr family \eqref{Kerr} (respectively the Reissner--Nordstr\"{o}m family \eqref{RN}) corresponding to $a=0$ (respectively $e=0$). Similarly, the extremal Reissner--Nordstr\"{o}m ($|e|=M$) and Kerr ($|a|=M$) solutions have co-dimension one within their respective families. In view of this, it is reasonable to conjecture that black holes converging to an extremal  Reissner--Nordstr\"{o}m/Kerr solution (asymptotically extremal black holes)  or Schwarzschild (asymptotically Schwarzschildean black holes) have non-generic initial data within the moduli space of regular solutions, which is why they should not play a  major role  in a proof of Conjecture~\ref{SCC.conj}.
	
	\textbf{Asymptotically extremal black holes.} The recent constructions of Kehle--Unger \cite{KehleUnger,KehleUnger2}  suggest that  asymptotically extremal black holes indeed arise for a finite codimensional subset of asymptotically flat initial data, at least within spherically symmetric models. Spherically-symmetric asymptotically extremal black hole solutions of \eqref{EMCSF} (with a small charge) were studied by Gajic--Luk \cite{gajicluk} (see also previous linear results by Gajic \cite{DejanlinintextI,DejanlinintextII}) showing that under a decay condition like \eqref{decay} on the event horizon, the spacetime admits a non-trivial Cauchy horizon $\mathcal{CH}^+$ (like in Theorem~\ref{CH.thm.SS}) and the metric is $C^0\cap H^1_{loc}$-extendible across the Cauchy horizon (in contrast to Theorem~\ref{CH.instab.thm.SS}). This shows that one can extend the metric as a weak solution to the Einstein--Maxwell-charged-scalar-field system of equations (recall the discussion in Statement~\ref{H1} of Section~\ref{subsection.MGHD}). An analogous statement for asymptotically extremal solutions  of the Einstein vacuum equations (without any symmetry assumptions) remains open.
	
	\textbf{Asymptotically Schwarzschildean black holes.} Within spherical symmetry,  black hole solutions of \eqref{ESF} with localized initial data are all asymptotically Schwarzschildean (this follows from \cite{PriceLaw}) and the MGHD  features a spacelike singularity $\mathcal{S}=\{r=0\}$, as discussed in Section~\ref{subsection.christodoulou}.
	For spherically-symmetric models with charge, in which asymptotically Schwarzschildean black holes are non-generic, it is shown in the author's \cite{MoiThesis} that asymptotically Schwarzschildean spherically-symmetric solutions of \eqref{EMCSF} have no Cauchy horizon ($\mathcal{CH}^+=\emptyset$) and the MGHD terminal boundary is foliated by spheres of area-radius $r=0$.
	For the Einstein vacuum equations, the result of Dafermos--Holzegel--Rodnianski--Taylor \cite{SchwarzschildStab} shows  that asymptotically Schwarzschildean black holes arise from a co-dimension three subset of initial data close to Schwarzschild. For these asymptotically Schwarzschildean black holes, the spacetime supremum of Kretschmann  curvature blows up \cite{SchwarzschildStab,KerrStab} in the  black hole interior, which  suggests that the MGHD is future-inextendible. 
	However, the causal  nature (null or spacelike) of the singularity, and quantitative estimates in its vicinity remain open, even within spherical symmetry. 
	
	\subsection{Spatially-compact spacetimes, cosmological analogues of Cosmic Censorship}\hskip 3 mm\textbf{Hawking singularity theorem and timelike incompleteness.} Spacetimes admitting a \emph{compact Cauchy surface} are particularly relevant to Cosmology as models of a finite universe and are subjected to the celebrated Hawking singularity  theorem  \cite{Hawking:sing} (compare to Theorem~\ref{penrose.sing.thm} which is, in contrast, more suitable to the context of gravitational collapse).	\begin{Theorem}[\cite{Hawking:sing}]\label{Hawking.sing.thm}If  $(\mathcal{M},g)$  
		admits a compact				Cauchy hypersurface with  positive mean-curvature
		and satisfies the  energy condition \eqref{time.energy}, then it is
		future-timelike geodesically incomplete.
	\end{Theorem} Similarly  to Theorem~\ref{penrose.sing.thm}, however, Theorem~\ref{Hawking.sing.thm} does not provide any information on the cause of geodesic incompleteness or the singular nature of the MGHD terminal boundary.
	
	\textbf{The rigidity of compact Cauchy horizons.} As in the gravitational collapse case, Cauchy horizons\footnote{As an example, the Taub--NUT spacetime has both a compact Cauchy surface and an extendible Cauchy horizon \cite{Hawking}.} can be a cause for  MGHD extendibility, and thus constitute the main obstruction to Strong Cosmic Censorship. A possible proof strategy of Strong Cosmic Censorship for  spatially-compact spacetimes  is to show that for solutions to the Einstein vacuum equations, any \emph{compact Cauchy horizon  is a Killing horizon}, as conjectured by Isenberg--Moncrief \cite{CH.conj}. Such ridigity results show that Cauchy horizons only exist in spacetimes with an extra symmetry and thus are non-generic. See \cite{CH.rigid2,CH.rigid} for a partial resolution   of the Isenberg--Moncrief problem.
	
	\textbf{Strong Cosmic Censorship in vacuum under Gowdy-symmetry.} Strong Cosmic Censorship was studied for a class of spacetimes with a compact spatial topology $\mathbb{T}^3$,  $\mathbb{S}^3$ or $\mathbb{S}^2\times \mathbb{S}^1$ admitting a two-dimensional isometry group $\mathbb{U}(1)\times \mathbb{U}(1)$ with spacelike orbits, under an additional two-surface orthogonality condition (Gowdy-symmetry, see \cite{Gowdy,Gowdy2,Ringstrom.review}). The Gowdy polarized subcase is defined as the instance in which the two independent Killing vector fields generated by the isometry group are orthogonal. In view of Theorem~\ref{Hawking.sing.thm}, one may expect these spacetimes to feature a spacelike singularity at $\{t=0\}$.  A version of Strong Cosmic Censorship within polarized-Gowdy solutions of the Einstein vacuum equations was proven by Chrusciel--Isenberg--Moncrief \cite{CIMSCC} for any $\mathbb{T}^3$,  $\mathbb{S}^3$ or $\mathbb{S}^2\times \mathbb{S}^1$ topology. Strong Cosmic Censorship within the more general unpolarized-Gowdy class was subsequently proved by Ringstr\"{o}m \cite{RingstromSCC,RingstromSCC2} in the $\mathbb{T}^3$ topology. Both unpolarized and polarized results proceed proving the blow-up of the Kretschmann scalar along incomplete geodesics,  employing  asymptotic expansions in the vicinity of the  singularity at $\{t=0\}$. 
	The unpolarized case is significantly more involved on a technical level due to the presence of so-called \emph{spikes} \cite{Ringstrom.review,RingstromSCC} which occur when the limit of some renormalized metric components (more precisely, their time-derivative)   as $t \rightarrow 0 $ is a discontinuous function on $\{t=0\}$. A key element of the proof in \cite{RingstromSCC,RingstromSCC2} is to construct  a subset of regular initial data $\mathcal{G}$ for which only finitely-many spikes occur in evolution  and such that  $\mathcal{G}$ is both dense in a smooth topology and open in a (possibly weaker) regular topology (this is the definition of $\mathcal{G}$ being a ``generic set'').
	
	\textbf{Strong Cosmic Censorship with collisionless matter under surface symmetry.} Dafermos--Rendall \cite{DR1,DR2,DR3} and Smulevici \cite{JacquesT2}  considered solutions of the Einstein--Vlasov equations for surface-symmetric spacetimes with topology $\mathbb{S}^1 \times \Sigma$, where $\Sigma$ is a 2-dimensional compact surface of constant curvature $k \in \{-1,0,1\}$ (note that the Gowdy symmetry class with  $\mathbb{S}^2\times \mathbb{S}^1$ topology is a subclass of the  $k=1$ case). They established the $C^2$ inextendibility of the MGHD (both for  vanishing  $\Lambda=0$ and  positive $\Lambda>0$ cosmological constant), thus proving a version of Strong Cosmic Censorship for collisionless matter within the surface-symmetry class. It is remarkable that, contrary to Gowdy-symmetric vacuum spacetimes,  precise asymptotics near the singularity are not essential in the case of collisionless matter considered in \cite{DR1,DR2,DR3,JacquesT2} and inextendibility  instead follows from softer geometric arguments, in particular the extension principle of \cite{DR2}.

	\textbf{Strong Cosmic Censorship  in the vicinity of Kasner spacetimes.} Recall the Kasner metric \eqref{gK} from Section~\ref{subsection.hairy} is an anisotropic spatially-homogeneous spacetime with compact spatial topology $\mathbb{T}^3$ and a spacelike singularity at $\{t=0\}$. Its perturbations for the Einstein vacuum equations may have very intricate dynamics (BKL heuristics \cite{BKL1,BKL2,mixmaster}). However, the dynamics is expected to admit stable fixed points represented by Kasner metrics of the form \eqref{gK} either    the presence of a scalar field (i.e., for solutions of \eqref{Einstein}, \eqref{ESF}) or within $\mathbb{U}(1)$-polarized  solutions of the Einstein vacuum equations.  The work of Fournodavlos--Rodnianski--Speck \cite{FournodavlosRodnianskiSpeck} proves the stability of such solutions, and  one can thus infer a version of Strong Cosmic Censorship  in a vicinity of these Kasner spacetimes for the Einstein-scalar-field equations (or the vacuum Einstein equations in $\mathbb{U}(1)$-polarized  symmetry).  The results of \cite{FournodavlosRodnianskiSpeck} were recently generalized by Groeninger--Peterson--Ringstr\"{o}m   \cite{Ringstrom23} proving the stability of a larger class of so-called ``quiescent singularities'' that contains the class considered in \cite{FournodavlosRodnianskiSpeck}. Thus, an analogous statement of Strong Cosmic Censorship can be obtained in the vicinity of the larger class of initial data considered in \cite{Ringstrom23}.

	\subsection{Strong Cosmic Censorship in higher-dimensional gravity}

	\textbf{\hskip 4 mm Higher-dimensional black hole stability.} There   are higher-dimensional analogues of  the Schwarzschild  \eqref{schw} and Kerr \eqref{Kerr} black holes, so-called Myers--Perry black holes \cite{MyersPerry}. The techniques used for black hole exterior stability in $(3+1)$-dimensions  discussed in Section~\ref{section.exterior} do not  apply to the higher-dimensional case; see, however, \cite{Collingbourne1,Collingbourne2,MetcalfeMyersPerry,HKWhigher,Volkerrp} for  interesting recent progress.

	\textbf{Extra stationary black holes: black rings.} The Schwarzschild  and Myers--Perry solutions are not the only stationary asymptotically-flat higher-dimensional black holes in view of the celebrated black ring solutions \cite{blackrings}, which possess an event horizon with $\mathbb{S}^2 \times\mathbb{S}^1$ topology. However, the black ring is subjected to the celebrated Gregory-Laflamme instability \cite{GL} taking the form of exponentially growing modes for \eqref{wave},  see \cite{GLproof} for a proof of this instability in $(4+1)$ dimensions. Additionally, stable trapping takes place for \eqref{wave} if $g$ is a black ring, leading to a slow logarithmic decay-in-time \cite{Benomio}, a potential source of further (nonlinear) instabilities.
	
	\textbf{Open problem.}	To summarize, even the conjectured endstate of higher-dimensional gravitational collapse appears to be more intricate than its $(3+1)$-dimensional counterpart; thus, an hypothetical higher-dimensional analogue of Conjecture~\ref{SCC.conj} remains completely open at present. 
	\subsection{Non-vanishing cosmological constant}
	
	\hskip 4 mm	\textbf{Positive cosmological constant.} The Schwarzschild-de-Sitter and Kerr-de-Sitter  (respectively Reissner--Nordstr\"{o}m-de-Sitter) spacetimes are black hole solutions of the Einstein vacuum (respectively Einstein--Maxwell) equations with a positive cosmological constant $\Lambda>0$. Their Cauchy surface (for a region inside the  cosmological horizon, thus excluding the cosmological region, see e.g.\ \cite{claylecturenotes}) is compact with the topology of $\mathbb{S}^2\times \mathbb{S}^1$, thus the de-Sitter black holes are not asymptotically flat (their interior region, however, is similar to their $\Lambda=0$ counterparts). Consequently, solutions of the wave  equation \eqref{wave}  do not decay at an inverse-polynomial rate (in the style of \eqref{decay})  on the event horizon, but instead at an exponential rate \cite{Fanglinear,Mavrogiannislinear,dyatlovenergydecay,Vasymethod}, see also \cite{HVQLW,Mavrogiannisnonlinear,HVSL} for such results  on nonlinear wave equations. At the level of the Einstein-vacuum equations, the stability of the Kerr-de-Sitter black holes (with $|a|\ll M$) at an exponential rate was proven in the celebrated work of Hintz--Vasy \cite{Andras}, reproved in \cite{Fangnonlinear}. We note that Theorem~\ref{CH.thm.vacuum} only requires decay upper bounds on the event horizon (compatible with an exponential rate) and thus applies mutatis mutandis to small perturbations of Kerr-de-Sitter black holes with $|a|\ll M$, which therefore possess a $C^0$-extendible Cauchy horizon. \\However, due to the fast exponential decay in the $\Lambda>0$ case, the singular character of the Cauchy horizon, and thus the validity of the $H^1$ formulation  of Strong Cosmic Censorship appears in doubt (see \cite{hintzvasyscc} and the appendix of \cite{nospacelike}).					Heuristics and numerical works on the wave equation \eqref{wave} indeed indicate that  solutions with smooth and localized initial data are actually   $H^1_{loc}$-bounded at the Cauchy horizon of a Reissner--Nordstr\"{o}m-de-Sitter hole  \cite{SCCheuristicsRN} or Kerr--Newman-de-Sitter \cite{SCCheuristicsKerrN} for sub-extremal parameters sufficiently close to extremality  (in contrast to the combined result of Theorem~\ref{Pricelaw.thm} and Theorem~\ref{linear.int.thm}  in the $\Lambda=0$ case); however, on a  Kerr-de-Sitter\footnote{The numerics of \cite{SCCheuristicsRN,SCCheuristicsKerr,SCCheuristicsKerrN} thus  surprisingly suggest  that Strong  Cosmic Censorship (at the $H^1$ level, i.e., Statement~\ref{H1} of Section~\ref{subsection.MGHD}) for $\Lambda>0$		
		may be violated in the presence of charge (Maxwell field), but respected in its absence.} black hole, solutions to \eqref{wave} seem to blow-up in  $H^1$ norm at the Cauchy horizon
	\cite{SCCheuristicsKerr}.		\\
	At the nonlinear level, spherically-symmetric Einstein--Maxwell-(charged)-scalar-field solutions of \eqref{EMCSF} approaching a (sub-extremal) Reissner--Nordstr\"{o}m-de-Sitter black hole  were studied in \cite{Flavio,Costa1,Costa2}.  Under the assumption of exponential decay of the scalar field on the event horizon, the stability of the Cauchy horizon was obtained analogously to Theorem~\ref{CH.thm.SS}. For a sufficiently rapidly-decaying scalar field on the event horizon, however, the Cauchy horizon is not weakly singular  and the Hawking mass remains bounded in this case (in complete contrast to Theorem~\ref{CH.instab.thm.SS} obtained in the $\Lambda=0$ case). Whether or not generic spherically-symmetric solutions decay sufficiently fast on the event horizon for this scenario to materialize remains open, however.
	
	\textbf{Negative cosmological constant.} On the Schwarzschild/Kerr-anti-de-Sitter black hole (solutions of the vacuum Einstein equations with $\Lambda<0$   analogous to \eqref{schw} and \eqref{Kerr}), solutions of the wave equation  \eqref{wave} decay at a slow inverse-logarithmic rate  \cite{quasimodeads,Gannot}.  On a Reissner--Nordstr\"{o}m-anti-de-Sitter black hole, a recent result of Zheng \cite{Weihao1} shows the existence of non-trivial stationary solutions of \eqref{wave} (see also previous heuristics works  \cite{holorigin2,Diashairy} in the context of AdS-CFT) and exponentially-growing modes (the former also being associated to asymptotically-AdS hairy black holes \cite{Weihao2}). Excluding these, Kehle  proved that solutions to \eqref{wave} are (point-wise) bounded at the Reissner--Nordstr\"{o}m-anti-de-Sitter Cauchy horizon \cite{Kehle2019}, although generically their $H^1$ norm blows-up \cite{Kehle2021}. On a Kerr-anti-de-Sitter black hole, in contrast, Kehle surprisingly showed   \cite{Kehle2020} that solutions of \eqref{wave} blow up point-wise at the Cauchy horizon, for black hole parameters $(M,|a|,\Lambda)$ belonging to a Baire-generic subset (of zero Lebesgue measure) of $\mathbb{R}_{+}^3$.  The exact repercussions of these instabilities on the Einstein-vacuum  dynamics, however, are unclear in view of the conjectured nonlinear instability of the Kerr-anti-de-Sitter exterior spacetime \cite{Figueras}.

	\section{Summary and future directions}
	This review explores the burgeoning field of Strong Cosmic Censorship research, which has developed its own momentum within the mathematics landscape of General Relativity, while remaining faithful to Penrose's original spirit.
	The understanding of  its key features  were  indeed made possible thanks to advancements in the theory of quasilinear waves, started in the 1990s, which drove immense progress in the study of spacetime dynamics in General Relativity.
	
	We highlight a few of these essential realizations which are all connected to breakthroughs in the understanding of black hole dynamics, towards a resolution of Problem~\ref{BH.problem}:   \begin{itemize}
		\item The Kerr Cauchy horizon is stable against dynamical perturbations, thus the singularity inside a generic black hole arising in gravitational collapse is not entirely-spacelike.
		\item The blue-shift instability, a linear mechanism on a stationary black hole with a smooth
		Cauchy horizon, does not destroy the Cauchy horizon when backreaction is included, but
		instead makes it weakly singular (proven only in spherical symmetry at present).
		\item A black hole formed in gravitational collapse  features a  spacelike singularity $\mathcal{S}$ that differs significantly from the singularity inside Schwarzschild/Oppenheimer--Snyder spacetimes and whose endpoint intersects  the null Cauchy horizon $\mathcal{CH}^+$  (proven only in the  charged   gravitational collapse of a spherically-symmetric scalar field at present).
	\end{itemize} 
	We also emphasize the remarkable  developments in the understanding of spacelike singularities dynamics. While the mathematical confirmation of the mixmaster/BKL heuristics for the Einstein vacuum equations remains an open problem, the important progress positions us well to tackle the challenges spacelike singularities pose within the resolution of Strong Cosmic Censorship.

	Moreover, and despite the independence of the Weak and Strong Cosmic Censorship conjectures highlighted in Section~\ref{subsection.independent}, it is clear that a complete resolution of Conjecture~\ref{SCC.conj} will require  comprehensive knowledge of the dynamics of (locally) naked singularities  in gravitational collapse, most notably their non-generic character: in other words, it is likely that a proof of  Weak Cosmic Censorship will provide essential insights to the proof of Conjecture~\ref{SCC.conj}.  Even a complete understanding of all the above phenomena without  any  symmetry assumption, however, would still not be sufficient to prove Conjecture~\ref{SCC.conj}, because it is tied to the comprehension of \emph{every generic endstate of gravitational collapse}, and therefore the final state conjecture needs to be resolved first. Despite this obstacle, progress on the mathematical program and partial steps towards Conjecture~\ref{SCC.conj} outlined in this review would have significant ramifications for understanding black hole dynamics and the field of General Relativity as a whole. 
	
	Among future directions, the study of multiple-black-holes dynamics, and whether they respect Strong Cosmic Censorship will be essential. Beyond the Einstein equations in vacuum, incorporating realistic astrophysical fluids, though presenting its own mathematical  challenges, holds immense potential for a more comprehensive physical understanding. As a steppingstone, studying a massive (Klein--Gordon) scalar field will already reveal a drastically different picture of Strong Cosmic Censorship compared to the simpler (electro)-vacuum case. 
	
	Strong Cosmic Censorship research stands as a vibrant frontier, pushing the boundaries of both General Relativity and mathematical analysis. Significant progress has been made, yet captivating challenges remain. As we delve deeper into this captivating domain, we stand to unlock a wealth of knowledge about the nature of gravity and the enigmatic realm of black holes.

	\section*{Acknowledgments}
	
	The author  acknowledges the support  of NSF Grant DMS-2247376. He also 	 thanks Hayd\'{e}e Pacheco  for the figures in this article, and Jonathan Luk for useful comments on the manuscript.

	\bibliographystyle{unsrt}

	\bibliography{bibliography}

\end{document}